\pdfoutput=1
\documentclass[aps,prl,twocolumn,superscriptaddress,showpacs,floatfix,preprintnumbers,nofootinbib]{revtex4-2}

\usepackage[colorlinks,allcolors=blue]{hyperref}
\usepackage{amsmath, amssymb, amsfonts}
\usepackage{graphicx}
\usepackage{bm}
\usepackage[normalem]{ulem}
\usepackage{color}
\usepackage{xcolor}
\usepackage{comment}
\usepackage{multirow}
\usepackage{color}
\usepackage{csquotes}
\usepackage{longtable}
\usepackage{scalerel}
\usepackage{siunitx}
\usepackage[section]{placeins}
\usepackage{wrapfig}
\usepackage{pgf}

\newcommand{\be}{\begin{equation}}
\newcommand{\ee}{\end{equation}}
\newcommand{\Tr}{\ensuremath{\mathrm{Tr}}}

\begin{document}

\preprint{TUM-EFT 204/26, TTK-26-24}
\title{Inclusive P-wave Quarkonium Decay Widths from Lattice QCD and pNRQCD}

\author{Nora Brambilla}
\email{nora.brambilla@ph.tum.de}
\affiliation{Technical University of Munich,\\
TUM School of Natural Sciences, Physics
Department,\\   
James-Franck-Str.~1, 85748 Garching, Germany.}
\affiliation{Technical University of Munich, Institute for Advanced Study, \\ 
Lichtenbergstrasse 2 a, 85748 Garching, Germany.}
\affiliation{Technical University of Munich, Munich Data Science Institute, \\ 
Walther-von-Dyck-Strasse 10, 85748 Garching, Germany.}

\author{Viljami Leino}
\email{leino@qtc.sdu.dk}
\affiliation{Quantum Theory Center ($\hslash$QTC) at IMADA \& D-IAS, University of Southern Denmark, \\ Campusvej 55, DK-5230 Odense M, Denmark}

\author{Julian Mayer-Steudte}
\email{julian.mayer-steudte@tum.de}
\affiliation{Technical University of Munich,\\
TUM School of Natural Sciences, Physics
Department,\\   
James-Franck-Str.~1, 85748 Garching, Germany.}
\affiliation{Technical University of Munich, Munich Data Science Institute, \\ 
Walther-von-Dyck-Strasse 10, 85748 Garching, Germany.}

\author{Panayiotis Panayiotou}
\email{panayiotis.panayiotou@tum.de}
\affiliation{Technical University of Munich,\\
TUM School of Natural Sciences, Physics
Department,\\   
James-Franck-Str.~1, 85748 Garching, Germany.}

\author{Andrea~Shindler}
\email{shindler@physik.rwth-aachen.de}
\affiliation{Institute for Theoretical Particle Physics and Cosmology, TTK, RWTH Aachen University, Sommerfeldstr. 16, Aachen, 52074, Germany}
\affiliation{Nuclear Science Division, Lawrence Berkeley National Laboratory, Berkeley, CA 94720, USA}
\affiliation{Department of Physics, University of California, Berkeley, CA 94720, USA}

\author{Antonio Vairo}
\email{antonio.vairo@tum.de}
\affiliation{Technical University of Munich,\\
TUM School of Natural Sciences, Physics
Department,\\   
James-Franck-Str.~1, 85748 Garching, Germany.}

\author{Xiang-Peng Wang}
\email{xpwang@ccnu.edu.cn}
\affiliation{Institute of Particle Physics and Key Laboratory of Quark and Lepton Physics (MOE),\\
Central China Normal University, Wuhan, Hubei 430079, China}

\newcommand{\msbar}{\ensuremath{\overline{\mathrm{MS}}}}

\date{\today}

\begin{abstract}
Inclusive hadronic decay widths remain a long-standing challenge for first-principles QCD. We present a framework combining lattice QCD with strongly-coupled potential nonrelativistic QCD (pNRQCD) to compute inclusive P-wave heavy quarkonium decays to light hadrons. At leading order in the velocity expansion, all nonperturbative effects, apart from the square of the derivative of the wavefunction at the origin, are encoded in a single universal moment of the two-point chromoelectric correlator,
which we determine for the first time from a quenched lattice QCD calculation matched to $\msbar$ via the gradient flow. Combined with perturbative short-distance coefficients and the square of the derivative of the wavefunction at the origin, our result reproduces the observed $\chi_{cJ}(1P)$ widths and, at the same time, provides predictions for the unmeasured $\chi_{bJ}(nP)$ widths. The framework extends naturally to inclusive decays and production of ordinary and exotic hadrons.
\end{abstract}

\maketitle

\paragraph{\textbf{Introduction.}}  
Computing inclusive hadronic decay widths from first principles is a
long-standing challenge for QCD. Lattice methods address them only
indirectly, through recently developed spectral-density and inclusive-rate
techniques~\cite{Harris:2016usb,Meyer:2017ydp,Hashimoto:2017wqo,Hansen:2019idp,Bailas:2020qmv,Gambino:2020crt,Gambino:2022dvu,Buzzicotti:2023qdv,Bruno:2024fqc,DelDebbio:2024lwm,ExtendedTwistedMass:2024myu,Lupo:2026vdj},
while perturbative factorization expresses the widths in terms of nonperturbative matrix elements, which are typically supplied by models or the very data that the calculation aims to predict. 
In this work, we present a method that closes this gap by combining lattice QCD with potential nonrelativistic QCD (pNRQCD), and apply it to inclusive P-wave quarkonium decays into light hadrons.

Quarkonia are nonrelativistic systems characterized by a hierarchy of
scales $m_Q\gg m_Qv\gg m_Q v^2, \Lambda_{\mathrm{QCD}}$, with $ v\ll 1$ the relative velocity and $m_Q$ the mass  of the heavy quark of flavor $Q$. 
Integrating out the hard scale $m_Q$ yields NRQCD \cite{Caswell:1985ui,Bodwin:1994jh,Brambilla:2004jw}, in which the inclusive decay rate of P-wave quarkonia into light hadrons 
reads at leading order in $v$ \cite{Bodwin:1992ye,Bodwin:1994jh}:
\begin{align}\label{eq:NRQCDfact}
    \Gamma\left(\chi_{QJ}\to \mathrm{LH}\right)&=\frac{2}{m_Q^2}\left( \frac{3N_c}{2\pi m_Q^2}\mathrm{Im}f_{1}\left(^{3}P_{J}\right)\left|R^{\prime}_{\chi_Q}\left(0\right)\right|^{2}\right.\nonumber\\
    &\hspace{-1.5cm} +\mathrm{Im}f_{8}\left(^{3}S_{1} \right)\left\langle \chi_{QJ}\right|\psi^{\dagger}T^{a}\chi\chi^{\dagger}T^{a}\psi\left|\chi_{QJ}\right\rangle \Bigg);
\end{align}
$N_c=3$ is the number of colors,  $\left\langle\chi_{QJ}\right|\psi^{\dagger}T^{a}\chi\chi^{\dagger}T^{a}\psi\left|\chi_{QJ}\right\rangle $ is a color-octet long-distance matrix element (LDME), $\left|R^{\prime}_{\chi_Q}\left(0\right)\right|^{2}$ is the squared derivative of the radial P-wave quarkonium wavefunction at the origin, and $\mathrm{Im} f_{1}\left(^{3}P_{J}\right) $, $\mathrm{Im}f_{8}\left(^{3}S_{1}\right)$ are (the imaginary parts of) perturbative short-distance coefficients (SDCs), known to  $\mathcal{O}\left(\alpha_s^3\right)$~\cite{Petrelli:1997ge}.
The octet LDME is essential: it cancels the dependence on the factorization scale $\Lambda$  
that the singlet SDC develops at $\mathcal{O}\left(\alpha_s^3\right)$ \cite{Bodwin:1994jh},
\begin{align}\label{eq:SDClog}
    \mathrm{Im}f_{1}\left(^{3}P_{J}\right)\left(\Lambda\right)\sim -n_{f}\frac{2C_{F}}{9N_{c}}\alpha_{s}^{3}\,\mathrm{log}\left(\frac{\Lambda}{2m_Q}\right),
\end{align}
where $n_f$ is the number of massless flavors and $C_F = T_F(N_c^2-1)/N_c = 4/3$ for $T_F=1/2$.
The dependence on $\Lambda$ was left unsubtracted in early calculations~\cite{Barbieri:1981xz}. 
Within NRQCD, the octet LDME is a state-dependent nonperturbative parameter 
different and independent from the quarkonium wavefunction.
Direct lattice determinations were attempted in Refs.~\cite{Bodwin:1996tg,Bodwin:2001mk}
but faced several obstacles, including the inability to take the continuum limit in lattice NRQCD.
Integrating out the soft scale $m_Qv$  leads to
pNRQCD ~\cite{Pineda:1997bj,Brambilla:1999xf}. 
In the strongly coupled regime $m_Qv^2 \ll \Lambda_{\rm QCD}$ — the regime appropriate to
excited charmonium and bottomonium states below the open-flavor threshold — 
a striking simplification occurs: the octet LDMEs of all P-wave quarkonia of both flavors 
reduce to the product of the squared derivative of the radial 
wavefunction at the origin and a single, flavor-independent, moment of a gluonic correlator,$\mathcal{E}_3$~\cite{Brambilla:2001xy,Brambilla:2002nu,Brambilla:2020xod}:
\begin{align}\label{eq:octmat}
    \left\langle \chi_{QJ}\right|\psi^{\dagger}T^{a}\chi\chi^{\dagger}T^{a}\psi\left|\chi_{QJ}\right\rangle=\frac{2T_F}{9N_cm_Q^2}\frac{3N_c}{2\pi}\left|R^{\prime}_{\chi_Q}\left(0\right)\right|^{2}\mathcal{E}_3.
\end{align}
Hence, the ratio of the color octet LDME and the derivative of the wavefunction at the origin, which
in NRQCD is a state-by-state input, becomes in pNRQCD one universal number, $\mathcal{E}_3$, that 
can be computed in lattice QCD. Performing such a calculation is the goal of the present work.

The universal quantity
$\mathcal{E}_3$ is defined as~\cite{Brambilla:2001xy,Brambilla:2002nu,Brambilla:2020xod}
\begin{align}\label{eq:E3def}
    \mathcal{E}_{3}\left(\Lambda\right)=\frac{T_{F}}{N_{c}}\int_{0}^{\infty}\mathrm{d}t\,t^{3}\,\mathcal{E}(t) ,
\end{align}
with
\begin{align}
       \mathcal{E}(t)&\equiv \langle 0|gE^{i,a}(t)\Phi^{ab}(t,0)gE^{i,b}(0)|0\rangle,
\label{eq:defE}
\end{align}
where $E^{i,a} = \partial_0 A_i^a - \partial_i A_0^a - gf^{abc} A_0^b A_i^c$ 
is the chromoelectric field, which in \eqref{eq:defE} is computed at the spatial origin and at the times $t$ and $0$,
and  $\Phi^{ab} (t,0)$ is a temporal Wilson line in the adjoint representation, ensuring the gauge invariance of the correlator \eqref{eq:defE}. 
At short $t$, $\mathcal{E}(t)$ is known perturbatively through next-to-next-to-leading
order~\cite{Eidemuller:1997bb,Braun:2020ymy,Braun:2021cqe} in the $\msbar$ scheme. 
At large $t$ it is genuinely nonperturbative. 
The dependence of $\mathcal{E}_3$ on the scale $\Lambda$ arises from the short-time region of the integral, where $\mathcal{E}_3$ develops an ultraviolet (UV) logarithmic divergence regularized into a $\log\Lambda$ term. 
This $\log\Lambda$ term cancels 
against the infrared (IR) logarithm of the singlet SDC in Eq.~\eqref{eq:SDClog}, making the physical width in Eq.~\eqref{eq:NRQCDfact} finite and $\Lambda$-independent.

A previous determination from charmonium data using Eqs.~\eqref{eq:NRQCDfact} and \eqref{eq:octmat} gave $\mathcal{E}_3(1\,\mathrm{GeV}) = 2.05^{+0.94}_{-0.65}$ \cite{Brambilla:2020xod};
our aim here is to compute $\mathcal{E}_3$ on the lattice directly from Eqs.~\eqref{eq:E3def}--\eqref{eq:defE}. 
Two issues must be addressed. 
First, the cancellation between the IR logarithm of the singlet SDC and the UV behavior of $\mathcal{E}_3$ requires that the perturbative and lattice calculations are performed in 
the same renormalization scheme. 
Second, the chromoelectric correlator $\mathcal{E}(t)$ is noisy on the lattice and carries a power divergence associated with the temporal Wilson line, absent in dimensional regularization but generic to any cutoff scheme, which must be removed before the continuum limit is taken.
We address both issues by using the gradient flow as a common intermediate regulator: 
it simultaneously renormalizes the chromoelectric field, suppresses lattice noise and provides a well-defined framework for matching to the $\msbar$ scheme used in the perturbative calculation.

\paragraph{\textbf{Determining $\mathcal{E}_3$}} 
We split $\mathcal{E}_3$ into a perturbative short-time part, $\mathcal{E}_3^\mathrm{pt}$, and a nonperturbative long-time part, $\mathcal{E}_3^\mathrm{np}$, as
\begin{align}
    \mathcal{E}_3(\Lambda) &= \mathcal{E}_3^\mathrm{pt}(\Lambda,t_{\ast})+\mathcal{E}_3^\mathrm{np}(t_{\ast}),
    \label{eq:decomposition}
\end{align}
where $t_{\ast}$ defines the temporal scale separating the two regimes. 
The correlator in the integral $\mathcal{E}_3^\mathrm{np}(t_{\ast})$ is computed on the lattice in gradient flow.
The correlator in the integral $\mathcal{E}_3^\mathrm{pt}(\Lambda,t_{\ast})$ is computed perturbatively. 
The two calculations are matched at  $t_{\ast}$ in the $\msbar$ scheme, as detailed in the lattice section below.
The  perturbative contribution to $\mathcal{E}_3$ at leading order (LO) in $\alpha_s$ gives
\begin{align}
    \mathcal{E}_{3}^\mathrm{pt}\left(\Lambda,t_{\ast}\right)  
    & = \frac{T_{F}}{N_{c}}\int_{0}^{t_{\ast}}\mathrm{d}t\,t^{3}\,\mathcal{E}\left(t\right)\nonumber\\
    &\hspace{-10mm}=24C_F\frac{\alpha_s}{4\pi}\left(\log\left(\frac{e^{2\gamma_E}\Lambda^{2}t^2_\ast}{4}\right)-\frac{8}{3}\right)+\mathcal{O}\left(\alpha_s^2\right)\label{eq:E3_perturbative}.
\end{align}
For illustration we present  here only the LO result, while the final prediction for $\mathcal{E}_3$ is based on the  next-to-leading order (NLO) expression given in the Supplemental Material (SM).
The $ \mathrm{log} \,\Lambda$ UV divergence cancels against the IR logarithmic divergence in $\mathrm{Im}\,f_1\left(^3P_J\right)$. 
Likewise, the dependence on the 
cutoff $t_{\ast}$ cancels against the corresponding dependence in  $\mathcal{E}_{3}^\mathrm{np}$,
leaving a final prediction for the width that is 
independent of both $\Lambda$ and $t_\ast$.
The nonperturbative contribution to $\mathcal{E}_3$, 
\begin{align}\label{eq:E3latticedef}
    \mathcal{E}_{3}^\mathrm{np}(t_\ast)&=\frac{T_{F}}{N_{c}}\int_{t_{\ast}}^{\infty}\mathrm{d}t\,t^{3}\,\mathcal{E}(t),
\end{align}
is evaluated using the continuum, zero flow-time extrapolated, and $\msbar$-matched integral of the correlator $\mathcal{E}(t)$ obtained from a lattice calculation.

\paragraph{\textbf{Lattice calculation.}} We compute the chromoelectric correlator, $\mathcal{E}(t)$, on the lattice using the gradient flow method~\cite{Narayanan:2006rf,Luscher:2009eq,Luscher:2010iy}. This method introduces a new scale, the flow time $\tau_F$, which can be associated with a smoothing radius of order $\sqrt{8\tau_F}$. Additionally, a reference scale $t_0$~\cite{Luscher:2010iy} can be defined based on the action density measured at finite flow time. In contrast to the Sommer scale $r_0$~\cite{Sommer:1993ce}, which is defined from the static quark-antiquark force, $t_0$ can be measured with similar precision but with less effort. We calculate $t_0/a^2$ for each of our lattice ensembles ($a$ is the lattice spacing) and use those values to express dimensionful quantities and to perform the continuum limit. To convert to physical units, we use $\sqrt{8t_0}/r_0=0.943$~\cite{DallaBrida:2019wur} and $r_0=\SI{0.5}{fm}$.

At positive flow time, the gradient flow renders composite operators finite and provides an intermediate regulator for the power divergence associated with the Wilson line~\cite{Polyakov:1980ca}. Such a power divergence is absent in dimensional regularization, in particular in the $\overline{\mathrm{MS}}$ scheme. 
Therefore, a matching to the $\overline{\mathrm{MS}}$ scheme is required when using gradient flow.

The chromoelectric correlator on the lattice is defined as
\begin{align}
    G_{EE}(t) = -\langle 0| gE_{i,a}^\mathrm{latt}(t)\Phi_{ab}^\mathrm{latt}(t,0)gE_{i,b}^\mathrm{latt}(0)| 0 \rangle,
    \label{eq:GEE_corr}
\end{align} 
where $E_{i,a}^\mathrm{latt}$ stands for a lattice definition of the chromoelectric component $E^{i,a}$ (see SM), while $\Phi_{ab}^\mathrm{latt}(t,0)$ denotes the lattice equivalent of the temporal Wilson line $\Phi_{ab}(t,0)$. The correlator is measured at different lattice spacings $a$, each along a flow-time axis. At finite flow time, $\tau_F$, the gradient flow renders the correlator finite, regulating in particular the power divergence, so that the continuum limit at fixed $\tau_F$ is well defined. 
In practice, the continuum extrapolation is performed in a region of flow times satisfying $\sqrt{8\tau_F} \gtrsim a$, which ensures that cutoff effects are under control. This choice is supported by numerical studies and by the studies of the $E$-field  for $\sqrt{8\tau_F} \gtrsim a$ in Ref.~\cite{Brambilla:2023fsi}.

We take the continuum limit by performing a linear fit in $a^2/t_0$ at a fixed $t/\sqrt{t_0}$ and $\sqrt{8\tau_F/t_0}$. 
In contrast to the continuum limit, the zero flow-time limit
requires subtraction of power divergence and matching procedure. 
Following Ref.~\cite{Brambilla:2023vwm}, the matching formula from the gradient flow scheme to the $\msbar$ scheme is given by
\begin{align}
G_{EE}^{\msbar}(t)=    \lim_{\tau_F\to0}
 e^{\left(\frac{d}{\sqrt{8\tau_F}}+m_0^{\msbar}\right)t}\,
 Z_{EE}^{\msbar}\, 
 G_{EE}(\tau_F,t)
.
    \label{eq:matching}
\end{align}
The exponential factorizes the Wilson line power divergence and depends on the nonperturbative coefficient $d$ and the conversion constant 
$m_0^{\msbar}$ to $\msbar$; 
$Z_{EE}^{\msbar}$ is a matching coefficient needed to convert the flowed chromoelectric correlator, once power divergence subtracted, to the corresponding one in the $\msbar$ scheme.
In perturbation theory, it is found to be $Z_{EE}^{\msbar} = 1+ \mathcal{O}(\alpha_s^2)$~\cite{Brambilla:2023vwm}.

\begin{figure}
    \centering
    \includegraphics[width=\linewidth]{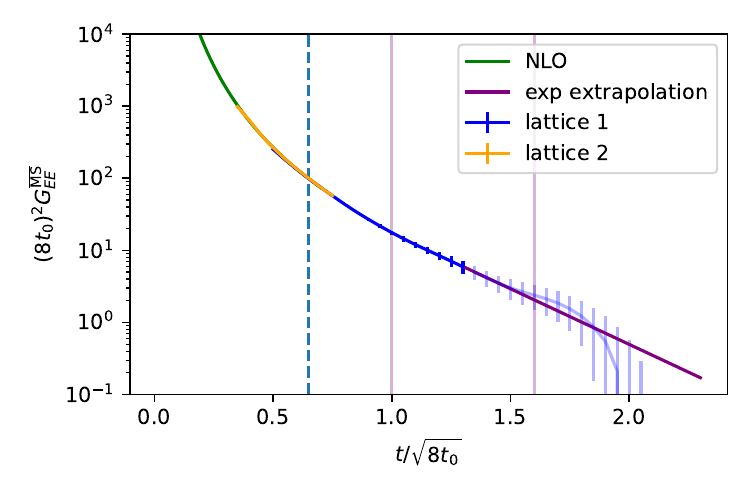}
    \caption{The matched correlator in $\overline{\mathrm{MS}}$ at a scale $\SI{1.17}{GeV}$ and $\alpha_s^{(n_f=0)}(\SI{1.17}{GeV})=0.26$. 
    ``NLO" is the curve based on the expression computed in~\cite{Eidemuller:1997bb}, ``lattice 1" and ``lattice 2" represent the lattice results from two different continuum limit approaches as discussed in the SM, and ``exp extrapolation" is the exponential function that is fitted to the data within the purple vertical lines to extrapolate 
    to large $t$.  
    The data points left of the dashed vertical line are used to extract $m_0^{\overline{\mathrm{MS}}}$.}
    \label{fig:Gr_msBar_matching}
\end{figure}

We determine $d$ from independent finite-temperature lattice calculations. 
In the continuum limit, we obtain nonperturbatively $d = 0.315(6)$ (see SM).
 Propagating the uncertainty of $d$ to the final result gives a contribution that is well below the statistical error and the other systematic uncertainties included in the analysis.  
 In this work, we perform the gradient flow to $\msbar$ matching at $\mathcal{O}(\alpha_s)$, that is, we set $Z_{EE}^{\msbar}=1$ and directly perform a linear extrapolation in $\tau_F$ for $\tau_F \to 0$ after subtracting the power divergence.  
 We determine $m_0^{\msbar}$ through comparing $G_{EE}^{\msbar}(t)$ in Eq.~(\ref{eq:matching}) with the NLO perturbative result of $\mathcal{E}^{\msbar}(t)$, and we get $m_0^{\msbar}\sqrt{8t_0} = 0.476(12)$. 
 The details of the determination of $m_0^{\overline{\mathrm{MS}}}$ and of the lattice calculation are given in the SM.
We set $\Lambda=2e^{-\gamma_E}/t_{\ast}\approx 1.17\,\mathrm{GeV}$ for $t_{\ast}\approx0.19\,\mathrm{fm}$. The final matched correlator in $\msbar$ is shown in Fig.~\ref{fig:Gr_msBar_matching}.

We are now in the position to compute the nonperturbative contribution to $\mathcal{E}_3$ using the lattice results extrapolated to the continuum and to vanishing flow time, with the matching described above,
\begin{align}
    \mathcal{E}^{\text{np}}_3(t_\ast)=\frac{T_F}{N_c}\int_{t_\ast}^\infty \mathrm{d}t ~t^3 G_{EE}^{\msbar}(t)\,.
    \label{eq:E3_lattice_part}
\end{align}
For the large-$t$ region, where the contribution to the integral is small, we estimate the tail using the ground-state dominance of the correlator $G_{EE}^{\msbar}(t)$. We numerically integrate this analytical expression with high precision using adaptive quadrature.
For the intermediate regime, we approximate the integral numerically with Simpson's rule.
 
Summing  $\mathcal{E}^{\text{np}}_3(t_\ast)$ with $\mathcal{E}^{\text{pt}}_3(\Lambda,t_\ast)$ evaluated at one loop, we obtain $\mathcal{E}_3(\Lambda)$ evaluated at the scale $\Lambda\approx 1.17\,\mathrm{GeV}$:
\begin{align}\label{eq: E3choice}
    \mathcal{E}_3\left(1.17\,\mathrm{GeV}\right)&=2.3\pm0.60^{\mathrm{stat.}}\pm0.15^{\mathrm{syst.}}\pm0.11^{\mathrm{h.o.}}.
\end{align}
The uncertainty associated with the choice of $t_{\ast}$ is small and is correlated with the statistical error of the lattice calculation.  For this reason, it is not included in our error budget.
We estimate effects due to higher order perturbative corrections in  $\mathcal{E}(t)$ by multiplying the perturbative expression by a prefactor $1\pm \alpha_s^2$ and repeating the matching procedure between the perturbative and lattice regions. This leads to a different value for $m_0^{\overline{\mathrm{MS}}}$, which in turn results in a different value for $\mathcal{E}_3$. The difference between the values of $\mathcal{E}_3$ defines our higher order (h.o.) error.

\paragraph{\textbf{Predictions for inclusive $\chi_c$ and $\chi_b$ decays.}}
Predictions for the inclusive decays require  that $\mathcal{E}_3$  is determined with dynamical fermion flavors. 
Using our quenched determination to estimate $\mathcal{E}_3$ in full QCD carries a systematic uncertainty.   
In~\cite{Bodwin:2001mk}, the authors evaluate this uncertainty to be about $14\%$.
In the present study, however, we do not include it in our error estimates. 
The fact that our predictions for the decay widths agree with data inside errors, see Table~\ref{tab:chibDW}, suggests that the systematic effect of unquenching is not larger than the considered errors.
Starting from Eq.~\eqref{eq: E3choice} and the leading-logarithmic renormalization group improvement of $\mathcal{E}_3$, 
\begin{align}\label{eq:RGE3}
\mathcal{E}_3\left(\Lambda\right)=\mathcal{E}_3\left(\Lambda^\prime\right)+\frac{24C_F}{\beta_0}\log\left(\frac{\alpha_s(\Lambda^\prime)}{\alpha_s(\Lambda)}\right),
\end{align}
 where $\beta_0=11C_A/3-4T_Fn_f/3$, we can  at least evolve $\mathcal{E}_3$ at the required scale with $n_f=3$ light fermions in the running for charmonium and $n_f=4$ for bottomonium. 
We set $m_Q$ to half the spin-averaged mass of the $\chi_{QJ}$ multiplet, $m_c = 1.76\,\mathrm{GeV}$ and $m_b = 4.95\,\mathrm{GeV}$, with $v^2$ corrections to this identification beyond our accuracy ($v^2\approx 0.3$ for charmonium and $v^2\approx 0.1$ for bottomonium).
We calculate $\left|R^{\prime}_{\chi_Q}\left(0\right)\right|^{2}$ by solving the Schrödinger equation with a Cornell potential of the form $V=-\gamma/r+\sigma r$, where $\gamma=0.434$ and $\sigma=0.198\,\mathrm{GeV^2}$~\cite{Brambilla:2024imu}.
We evaluate the SDCs, setting the hard renormalization scale to be the mass of the heavy quark $m_Q$.
From Eq.~\eqref{eq:RGE3}, we obtain $ \mathcal{E}_3\left(m_c\right)=3.17\pm0.60^{\mathrm{stat.}}\pm0.15^{\mathrm{syst.}}\pm0.11^{\mathrm{h.o.}}$
with $\alpha_s(m_c)=0.316$ and $\alpha_s(1.17{\rm GeV})=0.404$.
With this input, we calculate the decay rates of P-wave charmonia to light hadrons (LH) listed in Table~\ref{tab:chibDW}.
The truncation error that is associated with truncating the NRQCD factorization formula at the leading order in $v$ is estimated as $v^2$ times the central value. 
The error affecting the SDCs from neglecting higher-order contributions is estimated by multiplying 
their $\mathcal{O}(\alpha_s^3)$ term with $\pm\alpha_s$.
The results are consistent with experimental data within errors~\cite{ParticleDataGroup:2026aaa}. pNRQCD also allows to predict the decay rates of P-wave bottomonia to LH, which are not yet measured.
Using again Eq.~\eqref{eq:RGE3}, we obtain $\mathcal{E}_3\left(m_b\right)=4.78 \pm0.60^{\mathrm{stat.}}\pm0.15^{\mathrm{syst.}}\pm0.11^{\mathrm{h.o.}}$ with $\alpha_s\left(m_b\right)=0.212$. 
We then proceed as in the charmonium case. 
The second half of Table~\ref{tab:chibDW} reports our findings on P-wave bottomonium decays.
\begin{table}[h]
    \centering
   \begin{tabular}{|c|c|c|c|c|c|c|}
    \hline\hline
    $\chi_{QJ}$ & $\Gamma\left(\chi_{QJ}\to\mathrm{LH}\right)_{\mathrm{exp}}$ & $\Gamma\left(\chi_{QJ}\to\mathrm{LH}\right)$ & trunc. & SDC & $\mathcal{E}_3$ \\
    \hline\hline
    $\chi_{c0}(1P)$ &$12.2\pm0.5$ & $13.19 \pm 4.35 $ & $3.96$ & $1.79$ & $0.18$ \\       
    $\chi_{c1}(1P)$ &$0.552\pm0.041$ & $0.69 \pm 0.28$ & $0.21$ & $0.03$ & $0.18$  \\
    $\chi_{c2}(1P)$ &$1.60\pm0.06$ & $1.66 \pm 0.59$ & $0.50$ & $0.25$ & $0.18$ \\
    \hline\hline
    $\chi_{b0}(1P)$ & - & $1.11 \pm 0.14$ & $0.11$ & $0.08$ & $0.02$ \\       
    $\chi_{b1}(1P)$ & - & $0.13 \pm 0.02$ & $0.01$ & $0.01$ & $0.02$ \\
    $\chi_{b2}(1P)$ & - & $0.25 \pm 0.04$ & $0.03$ & $0.01$ & $0.02$ \\
    $\chi_{b0}(2P)$ & - & $1.42 \pm 0.17$ & $0.14$ & $0.10$ & $0.02$ \\       
    $\chi_{b1}(2P)$ & - & $0.17 \pm 0.03$ & $0.02$ & $0.01$ & $0.02$ \\
    $\chi_{b2}(2P)$ & - & $0.32 \pm 0.04$ & $0.03$ & $0.01$ & $0.02$ \\
     $\chi_{b0}(3P)$ & - & $1.64 \pm 0.20$ & $0.16$ & $0.11$ & $0.03$ \\       
    $\chi_{b1}(3P)$ & - & $0.19 \pm 0.04$ & $0.02$ & $0.01$ & $0.03$ \\
    $\chi_{b2}(3P)$ & - & $0.36 \pm 0.05$ & $0.04$ & $0.01$ & $0.03$ \\
    \hline\hline
    \end{tabular}    
    \caption{Our predictions for the decay widths of $\chi_{c}$ and $\chi_{b}$ to $\mathrm{LH}$ are denoted  $\Gamma\left(\chi_{QJ}\to\mathrm{LH}\right)$.
    We include detailed error determinations for each decay width.
    The total errors are obtained from the sums in quadrature of the single errors. 
    We have included errors that involve the truncation in the velocity expansion ($\mathrm{trunc.}$), errors from higher orders in the perturbative expansion of the short-distance coefficients ($\mathrm{SDC}$), 
    and errors in the determination of the gluonic correlator $\mathcal{E}_{3}$. 
    We also give the available experimental determinations of the decay widths, denoted  $\Gamma\left(\chi_{QJ}\to\mathrm{LH}\right)_{\mathrm{exp}}$~\cite{ParticleDataGroup:2026aaa}. 
    All quantities are in $\mathrm{MeV}$ units.}
    \label{tab:chibDW}
\end{table}

\begin{figure}[ht]
\centerline
{
\input{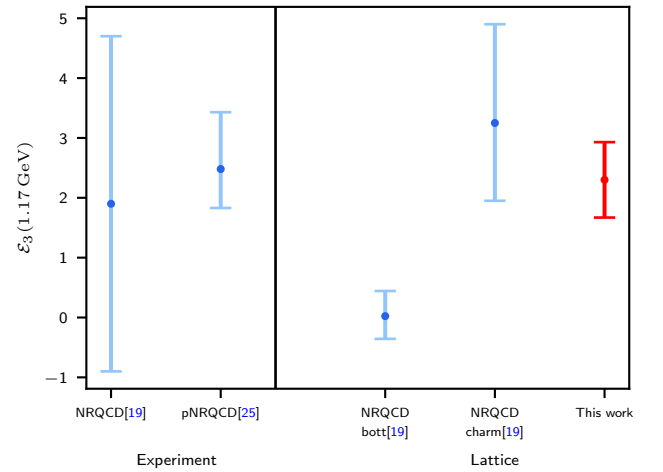}}
\caption{Comparison plot for the nonperturbative universal correlator $\mathcal{E}_3\left(1.17\,\mathrm{GeV}\right)$ including results from \cite{Bodwin:1996tg}, \cite{Brambilla:2020xod} and this work  (see text). 
The values obtained from experiments are run to a common scale $\Lambda=1.17\,\mathrm{GeV}$ using the appropriate number $n_f$ of active  massless flavors. 
 When comparing quenched lattice results, we run to the common scale with $n_f=0$.
\label{fig:comp}}
\end{figure}

\paragraph{\textbf{Discussion and Outlook.}}
In Fig.~\ref{fig:comp}, we compare our direct lattice determination of $\mathcal{E}_3$ at $\Lambda = 1.17\,\mathrm{GeV}$ with the extraction 
from charmonium data of the ratio of the NRQCD color octet LDME and the derivative of the wavefunction at the origin squared done in \cite{Bodwin:1996tg}, 
the data-driven pNRQCD extraction of Ref.~\cite{Brambilla:2020xod}, 
and the quenched lattice NRQCD computation of the ratio of the color octet LDME and the derivative of the wavefunction at the origin squared for bottomonium and charmonium done in \cite{Bodwin:1996tg}.
The reported values are obtained with completely different methods and inputs, therefore it is significant that almost all of them agree.
The only exception is the determination from the NRQCD bottomonium LDMEs of Ref.~\cite{Bodwin:1996tg}; 
it should be noted that this result also appears to violate the universality of the pNRQCD factorization.
Among the determinations that agree, the present one exhibits the smallest error. 

The methodology developed here addresses three obstacles that typically arise whenever inclusive observables are computed by combining lattice QCD with a nonrelativistic effective field theory:
interfacing perturbative and lattice calculations in a consistent renormalization scheme, regulating power divergences from Wilson-line operators via the gradient flow, and matching unambiguously to $\msbar$. 
These same obstacles appear in a large class of problems in QCD, and the present work provides a concrete template for tackling them. Several extensions follow naturally. 
The quenched approximation can be lifted by including dynamical fermions; 
higher-order corrections in $v^2$, which currently dominate the charmonium error budget, can be incorporated within the same pNRQCD framework. 
The method extends directly to S-wave quarkonium inclusive decays, where additional universal correlators 
appear \cite{Brambilla:2020xod}. 
A particularly direct extension is to inclusive quarkonium production: in pNRQCD, as the color-octet production LDMEs of P-wave quarkonia are controlled by a correlator that 
differs from $\mathcal{E}$ only by Wilson-line insertions, 
shares its one-loop renormalization-group evolution, 
and is accessible by the same lattice machinery developed here. 
Analogous correlators govern S-wave quarkonium production \cite{Brambilla:2022ayc, Brambilla:2022rjd} and the inclusive production of exotic states such as $\chi_{c1}(3872)$ and pentaquarks \cite{Lai:2025tpw,Brambilla:2026ujo}. 
Together, these constitute a small set of universal gluonic quantities from which a large body of inclusive quarkonium phenomenology can be reconstructed from first principles.
In summary, the considered method is, in its most general applications, of relevance
for inclusive observables at LHCb and Belle II, for the production of exotic hadrons that are central to the experimental program of the next decade, and for the determination of gluonic matrix elements entering transverse-momentum-dependent factorization in hadron structures of importance for the Electron-Ion Collider.
\newline

\begin{acknowledgments}
We would like to thank Damir Becirevic and Geoffrey T. Bodwin for useful discussions. 
The simulations were carried out on the computing facilities of the Computational Center for Particle and Astrophysics (C2PAP) in the project 
\emph{Calculation of finite $T$ QCD correlators} (pr83pu) and of the SuperMUC cluster at the Leibniz-Rechenzentrum (LRZ) in the project 
\emph{Static force and other operators with field strength tensor insertions} (pn49lo), 
both located in Munich (Germany).
The authors gratefully acknowledge the Gauss Centre for Supercomputing e.V. 
(\href{www.gauss-centre.eu}{www.gauss-centre.eu})
for funding this project by providing computing time on the GCS Supercomputer SuperMUC-NG 
at Leibniz Supercomputing Centre (\href{www.lrz.de}{www.lrz.de}). 
We acknowledge the DFG cluster of excellence ORIGINS funded by the Deutsche Forschungsgemeinschaft under Germany’s Excellence Strategy-EXC-2094-390783311. 
N.B. acknowledges the ERC Advanced Grant ERC-2023-ADG- Project EFT-XYZ.
N.B. and J.M.-S. acknowledge the DFG Grant No. BR 4058/5-1.
The work of V.L. is supported by the Carlsberg Foundation grant CF22-0922. 
The work of X.-P. W. is supported by the National Natural Science Foundation of China under Grant No. 12135006.
\end{acknowledgments}

\bibliography{references}

@article{Polyakov:1980ca,
    author = "Polyakov, Alexander M.",
    title = "{Gauge Fields as Rings of Glue}",
    doi = "10.1016/0550-3213(80)90507-6",
    journal = "Nucl. Phys. B",
    volume = "164",
    pages = "171--188",
    year = "1980"
}

@article{Lupo:2026vdj,
    author = "Lupo, Alessandro and Tantalo, Nazario",
    title = "{Extraction of spectral densities from lattice correlators: decoupling signal from noise}",
    eprint = "2605.14652",
    journal = "",
    archivePrefix = "arXiv",
    primaryClass = "hep-lat",
    month = "5",
    year = "2026"
}

@article{DelDebbio:2024lwm,
    author = "Del Debbio, Luigi and Lupo, Alessandro and Panero, Marco and Tantalo, Nazario",
    title = "{Bayesian solution to the inverse problem and its relation to Backus{\textendash}Gilbert methods}",
    eprint = "2409.04413",
    archivePrefix = "arXiv",
    primaryClass = "hep-lat",
    doi = "10.1140/epjc/s10052-025-13885-9",
    journal = "Eur. Phys. J. C",
    volume = "85",
    number = "2",
    pages = "185",
    year = "2025"
}

@article{Bruno:2024fqc,
    author = "Bruno, Mattia and Giusti, Leonardo and Saccardi, Matteo",
    title = "{Spectral densities from Euclidean lattice correlators via the Mellin transform}",
    eprint = "2407.04141",
    archivePrefix = "arXiv",
    primaryClass = "hep-lat",
    doi = "10.1103/PhysRevD.111.094515",
    journal = "Phys. Rev. D",
    volume = "111",
    number = "9",
    pages = "094515",
    year = "2025"
}

@article{Buzzicotti:2023qdv,
    author = "Buzzicotti, Michele and De Santis, Alessandro and Tantalo, Nazario",
    title = "{Teaching to extract spectral densities from lattice correlators to a broad audience of learning-machines}",
    eprint = "2307.00808",
    archivePrefix = "arXiv",
    primaryClass = "hep-lat",
    doi = "10.1140/epjc/s10052-024-12399-0",
    journal = "Eur. Phys. J. C",
    volume = "84",
    number = "1",
    pages = "32",
    year = "2024"
}

@article{Gambino:2020crt,
    author = "Gambino, Paolo and Hashimoto, Shoji",
    title = "{Inclusive Semileptonic Decays from Lattice QCD}",
    eprint = "2005.13730",
    archivePrefix = "arXiv",
    primaryClass = "hep-lat",
    reportNumber = "KEK-CP-376",
    doi = "10.1103/PhysRevLett.125.032001",
    journal = "Phys. Rev. Lett.",
    volume = "125",
    number = "3",
    pages = "032001",
    year = "2020"
}

@article{Gambino:2022dvu,
    author = {Gambino, Paolo and Hashimoto, Shoji and M{\"a}chler, Sandro and Panero, Marco and Sanfilippo, Francesco and Simula, Silvano and Smecca, Antonio and Tantalo, Nazario},
    title = "{Lattice QCD study of inclusive semileptonic decays of heavy mesons}",
    eprint = "2203.11762",
    archivePrefix = "arXiv",
    primaryClass = "hep-lat",
    doi = "10.1007/JHEP07(2022)083",
    journal = "JHEP",
    volume = "07",
    pages = "083",
    year = "2022"
}

@article{Bailas:2020qmv,
    author = "Bailas, Gabriela and Hashimoto, Shoji and Ishikawa, Tsutomu",
    title = "{Reconstruction of smeared spectral function from Euclidean correlation functions}",
    eprint = "2001.11779",
    archivePrefix = "arXiv",
    primaryClass = "hep-lat",
    reportNumber = "KEK-CP-374",
    doi = "10.1093/ptep/ptaa044",
    journal = "PTEP",
    volume = "2020",
    number = "4",
    pages = "043B07",
    year = "2020"
}

@article{Hashimoto:2017wqo,
    author = "Hashimoto, Shoji",
    title = "{Inclusive semi-leptonic B meson decay structure functions from lattice QCD}",
    eprint = "1703.01881",
    archivePrefix = "arXiv",
    primaryClass = "hep-lat",
    reportNumber = "KEK-CP-355",
    doi = "10.1093/ptep/ptx052",
    journal = "PTEP",
    volume = "2017",
    number = "5",
    pages = "053B03",
    year = "2017"
}

@article{Meyer:2017ydp,
    author = "Meyer, Harvey",
    title = "{Lattice QCD, Spectral Functions and Transport Coefficients}",
    doi = "10.22323/1.281.0364",
    journal = "PoS",
    volume = "INPC2016",
    pages = "364",
    year = "2017"
}

@article{Harris:2016usb,
    author = "Harris, Tim and Meyer, Harvey B. and Robaina, Daniel",
    title = "{A variational method for spectral functions}",
    eprint = "1611.02499",
    archivePrefix = "arXiv",
    primaryClass = "hep-lat",
    doi = "10.22323/1.256.0339",
    journal = "PoS",
    volume = "LATTICE2016",
    pages = "339",
    year = "2016"
}

@article{Hansen:2019idp,
    author = "Hansen, Martin and Lupo, Alessandro and Tantalo, Nazario",
    title = "{Extraction of spectral densities from lattice correlators}",
    eprint = "1903.06476",
    archivePrefix = "arXiv",
    primaryClass = "hep-lat",
    doi = "10.1103/PhysRevD.99.094508",
    journal = "Phys. Rev. D",
    volume = "99",
    number = "9",
    pages = "094508",
    year = "2019"
}

@article{ExtendedTwistedMass:2024myu,
    author = "Alexandrou, Constantia and others",
    collaboration = "Extended Twisted Mass",
    title = "{Inclusive Hadronic Decay Rate of the {\ensuremath{\tau}} Lepton from
Lattice QCD: The u{\textasciimacron}s Flavor Channel and the Cabibbo Angle}",
    eprint = "2403.05404",
    archivePrefix = "arXiv",
    primaryClass = "hep-lat",
    doi = "10.1103/PhysRevLett.132.261901",
    journal = "Phys. Rev. Lett.",
    volume = "132",
    number = "26",
    pages = "261901",
    year = "2024"
}

@article{Brambilla:2022ayc,
    author = "Brambilla, Nora and Chung, Hee Sok and Vairo, Antonio and Wang, Xiang-Peng",
    title = "{Inclusive production of J/{\ensuremath{\psi}}, {\ensuremath{\psi}}(2S), and {\ensuremath{\Upsilon}} states in pNRQCD}",
    eprint = "2210.17345",
    archivePrefix = "arXiv",
    primaryClass = "hep-ph",
    reportNumber = "TUM-EFT 170/22",
    doi = "10.1007/JHEP03(2023)242",
    journal = "JHEP",
    volume = "03",
    pages = "242",
    year = "2023"
}

@article{Barbieri:1981xz,
    author = "Barbieri, Riccardo and Caffo, Michele and Gatto, Raoul and Remiddi, E.",
    title = "{QCD corrections to P wave quarkonium decays}",
    reportNumber = "CERN-TH-3071",
    doi = "10.1016/0550-3213(81)90192-9",
    journal = "Nucl. Phys. B",
    volume = "192",
    pages = "61--65",
    year = "1981"
}

@article{Bodwin:1992ye,
    author = "Bodwin, Geoffrey T. and Braaten, Eric and Lepage, G. Peter",
    title = "{Rigorous QCD predictions for decays of P wave quarkonia}",
    eprint = "hep-lat/9205006",
    archivePrefix = "arXiv",
    reportNumber = "ANL-HEP-PR-92-30, NUHEP-TH-92-4",
    doi = "10.1103/PhysRevD.46.R1914",
    journal = "Phys. Rev. D",
    volume = "46",
    pages = "R1914--R1918",
    year = "1992"
}

@article{Bilson-Thompson:2002xlt,
    author = "Bilson-Thompson, Sundance O. and Leinweber, Derek B. and Williams, Anthony G.",
    title = "{Highly improved lattice field strength tensor}",
    eprint = "hep-lat/0203008",
    archivePrefix = "arXiv",
    reportNumber = "ADP-01-50-T482",
    doi = "10.1016/S0003-4916(03)00009-5",
    journal = "Annals Phys.",
    volume = "304",
    pages = "1--21",
    year = "2003"
}

@article{Brambilla:2002nu,
    author = "Brambilla, Nora and Eiras, Dolors and Pineda, Antonio and Soto, Joan and Vairo, Antonio",
    title = "{Inclusive decays of heavy quarkonium to light particles}",
    eprint = "hep-ph/0208019",
    archivePrefix = "arXiv",
    reportNumber = "CERN-TH-2002-179, IFUM-719-FT, UB-ECM-PF-02-15",
    doi = "10.1103/PhysRevD.67.034018",
    journal = "Phys. Rev. D",
    volume = "67",
    pages = "034018",
    year = "2003"
}

@article{Brambilla:2001xy,
    author = "Brambilla, Nora and Eiras, Dolors and Pineda, Antonio and Soto, Joan and Vairo, Antonio",
    title = "{New predictions for inclusive heavy quarkonium P wave decays}",
    eprint = "hep-ph/0109130",
    archivePrefix = "arXiv",
    reportNumber = "CERN-TH-2001-249, IFUM-692-FT, TTP-01-20, UB-ECM-PF-01-10",
    doi = "10.1103/PhysRevLett.88.012003",
    journal = "Phys. Rev. Lett.",
    volume = "88",
    pages = "012003",
    year = "2002"
}

@article{Brambilla:2020xod,
    author = {Brambilla, Nora and Chung, Hee Sok and M\"uller, Daniel and Vairo, Antonio},
    title = "{Decay and electromagnetic production of strongly coupled quarkonia in pNRQCD}",
    eprint = "2002.07462",
    archivePrefix = "arXiv",
    primaryClass = "hep-ph",
    reportNumber = "TUM-EFT 126/19",
    doi = "10.1007/JHEP04(2020)095",
    journal = "JHEP",
    volume = "04",
    pages = "095",
    year = "2020"
}

@article{Petrelli:1997ge,
    author = "Petrelli, Andrea and Cacciari, Matteo and Greco, Mario and Maltoni, Fabio and Mangano, Michelangelo L.",
    title = "{NLO production and decay of quarkonium}",
    eprint = "hep-ph/9707223",
    archivePrefix = "arXiv",
    reportNumber = "CERN-TH-97-142, DESY-97-090",
    doi = "10.1016/S0550-3213(97)00801-8",
    journal = "Nucl. Phys. B",
    volume = "514",
    pages = "245--309",
    year = "1998"
}

@article{Braun:2021cqe,
    author = "Braun, V. M. and Chetyrkin, K. G. and Kniehl, B. A.",
    title = "{Operator product expansion of the non-local gluon condensate}",
    eprint = "2103.09478",
    archivePrefix = "arXiv",
    primaryClass = "hep-ph",
    reportNumber = "Report-no: DESY 21-033, TTP21-005",
    doi = "10.1007/JHEP05(2021)231",
    journal = "JHEP",
    volume = "05",
    pages = "231",
    year = "2021"
}

@article{Braun:2020ymy,
    author = "Braun, V. M. and Chetyrkin, K. G. and Kniehl, B. A.",
    title = "{Renormalization of parton quasi-distributions beyond the leading order: spacelike vs. timelike}",
    eprint = "2004.01043",
    archivePrefix = "arXiv",
    primaryClass = "hep-ph",
    reportNumber = "DESY-20-060, Report-no: DESY 20-060, TTP20-015",
    doi = "10.1007/JHEP07(2020)161",
    journal = "JHEP",
    volume = "07",
    pages = "161",
    year = "2020"
}

@article{Eidemuller:1997bb,
    author = "Eidemüller, Markus and Jamin, Matthias",
    title = "{QCD field strength correlator at the next-to-leading order}",
    eprint = "hep-ph/9709419",
    archivePrefix = "arXiv",
    reportNumber = "HD-THEP-97-49",
    doi = "10.1016/S0370-2693(97)01352-X",
    journal = "Phys. Lett. B",
    volume = "416",
    pages = "415--420",
    year = "1998"
}

@article{Bodwin:1994jh,
    author = "Bodwin, Geoffrey T. and Braaten, Eric and Lepage, G. Peter",
    title = "{Rigorous QCD analysis of inclusive annihilation and production of heavy quarkonium}",
    eprint = "hep-ph/9407339",
    archivePrefix = "arXiv",
    reportNumber = "ANL-HEP-PR-94-24, FERMILAB-PUB-94-073-T, NUHEP-TH-94-5",
    doi = "10.1103/PhysRevD.55.5853",
    journal = "Phys. Rev. D",
    volume = "51",
    pages = "1125--1171",
    year = "1995",
    note = "[Erratum: Phys.Rev.D 55, 5853 (1997)]"
}

@article{Bodwin:2001mk,
    author = "Bodwin, Geoffrey T. and Sinclair, D. K. and Kim, S.",
    title = "{Bottomonium decay matrix elements from lattice QCD with two light quarks}",
    eprint = "hep-lat/0107011",
    archivePrefix = "arXiv",
    reportNumber = "ANL-HEP-PR-01-026",
    doi = "10.1103/PhysRevD.65.054504",
    journal = "Phys. Rev. D",
    volume = "65",
    pages = "054504",
    year = "2002"
}

@article{Bodwin:1996tg,
    author = "Bodwin, Geoffrey T. and Sinclair, D. K. and Kim, S.",
    title = "{Quarkonium decay matrix elements from quenched lattice QCD}",
    eprint = "hep-lat/9605023",
    archivePrefix = "arXiv",
    reportNumber = "ANL-HEP-PR-96-28",
    doi = "10.1103/PhysRevLett.77.2376",
    journal = "Phys. Rev. Lett.",
    volume = "77",
    pages = "2376--2379",
    year = "1996"
}

@article{Brambilla:2024imu,
    author = "Brambilla, Nora and Mohapatra, Abhishek and Scirpa, Tommaso and Vairo, Antonio",
    title = "{Nature of {\ensuremath{\chi}}c1(3872) and Tcc+(3875)}",
    eprint = "2411.14306",
    archivePrefix = "arXiv",
    primaryClass = "hep-ph",
    reportNumber = "TUM-EFT 193/24",
    doi = "10.1103/pdy7-hvg7",
    journal = "Phys. Rev. Lett.",
    volume = "135",
    number = "13",
    pages = "131902",
    year = "2025"
}

@article{Chetyrkin:2000yt,
    author = "Chetyrkin, K. G. and Kuhn, Johann H. and Steinhauser, M.",
    title = "{RunDec: A Mathematica package for running and decoupling of the strong coupling and quark masses}",
    eprint = "hep-ph/0004189",
    archivePrefix = "arXiv",
    reportNumber = "DESY-00-034, TTP-00-05",
    doi = "10.1016/S0010-4655(00)00155-7",
    journal = "Comput. Phys. Commun.",
    volume = "133",
    pages = "43--65",
    year = "2000"
}

@article{Brambilla:2004jw,
    author = "Brambilla, Nora and Pineda, Antonio and Soto, Joan and Vairo, Antonio",
    title = "{Effective Field Theories for Heavy Quarkonium}",
    eprint = "hep-ph/0410047",
    archivePrefix = "arXiv",
    reportNumber = "IFUM-805-FT, UB-ECM-PF-04-24",
    doi = "10.1103/RevModPhys.77.1423",
    journal = "Rev. Mod. Phys.",
    volume = "77",
    pages = "1423",
    year = "2005"
}

@article{Caswell:1985ui,
    author = "Caswell, W. E. and Lepage, G. P.",
    title = "{Effective Lagrangians for Bound State Problems in QED, QCD, and Other Field Theories}",
    reportNumber = "CLNS-85/641",
    doi = "10.1016/0370-2693(86)91297-9",
    journal = "Phys. Lett. B",
    volume = "167",
    pages = "437--442",
    year = "1986"
}

@article{Narayanan:2006rf,
    author = "Narayanan, R. and Neuberger, H.",
    title = "{Infinite N phase transitions in continuum Wilson loop operators}",
    eprint = "hep-th/0601210",
    archivePrefix = "arXiv",
    doi = "10.1088/1126-6708/2006/03/064",
    journal = "JHEP",
    volume = "03",
    pages = "064",
    year = "2006"
}

@article{Luscher:2009eq,
    author = "Lüscher, Martin",
    title = "{Trivializing maps, the Wilson flow and the HMC algorithm}",
    eprint = "0907.5491",
    archivePrefix = "arXiv",
    primaryClass = "hep-lat",
    reportNumber = "CERN-PH-TH-2009-118",
    doi = "10.1007/s00220-009-0953-7",
    journal = "Commun. Math. Phys.",
    volume = "293",
    pages = "899--919",
    year = "2010"
}

@article{Luscher:2010iy,
    author = {L{\"u}scher, Martin},
    title = "{Properties and uses of the Wilson flow in lattice QCD}",
    eprint = "1006.4518",
    archivePrefix = "arXiv",
    primaryClass = "hep-lat",
    reportNumber = "CERN-PH-TH-2010-143",
    doi = "10.1007/JHEP08(2010)071",
    journal = "JHEP",
    volume = "08",
    pages = "071",
    year = "2010",
    note = "[Erratum: JHEP 03, 092 (2014)]"
}

@article{Brambilla:2023fsi,
    author = "Brambilla, Nora and Leino, Viljami and Mayer-Steudte, Julian and Vairo, Antonio",
    title = "{Static force from generalized Wilson loops on the lattice using the gradient flow}",
    eprint = "2312.17231",
    archivePrefix = "arXiv",
    primaryClass = "hep-lat",
    reportNumber = "TUM-EFT 175/22, MITP-23-080",
    doi = "10.1103/PhysRevD.109.114517",
    journal = "Phys. Rev. D",
    volume = "109",
    number = "11",
    pages = "114517",
    year = "2024"
}

@article{Brambilla:2025cqy,
    author = "Brambilla, Nora and Datta, Saumen and Janer, Marc and Leino, Viljami and Mayer-Steudte, Julian and Petreczky, Peter and Vairo, Antonio",
    collaboration = "TUMQCD",
    title = "{Lattice study of correlators of chromoelectric fields for heavy quarkonium dynamics in the quark-gluon plasma}",
    eprint = "2505.16603",
    archivePrefix = "arXiv",
    primaryClass = "hep-lat",
    reportNumber = "TUM-EFT 189/24, MITP-24-077",
    doi = "10.1103/387k-mdl1",
    journal = "Phys. Rev. D",
    volume = "112",
    number = "7",
    pages = "074509",
    year = "2025"
}

@article{Francis:2015lha,
    author = "Francis, A. and Kaczmarek, O. and Laine, M. and Neuhaus, T. and Ohno, H.",
    title = "{Critical point and scale setting in SU(3) plasma: An update}",
    eprint = "1503.05652",
    archivePrefix = "arXiv",
    primaryClass = "hep-lat",
    doi = "10.1103/PhysRevD.91.096002",
    journal = "Phys. Rev. D",
    volume = "91",
    number = "9",
    pages = "096002",
    year = "2015"
}

@article{hoffman2014no,
  title={The No-U-Turn sampler: adaptively setting path lengths in Hamiltonian Monte Carlo.},
  author={Hoffman, Matthew D and Gelman, Andrew and others},
  journal={J. Mach. Learn. Res.},
  volume={15},
  number={1},
  pages={1593--1623},
  year={2014}
}

@article{Brambilla:2023vwm,
    author = "Brambilla, Nora and Wang, Xiang-Peng",
    title = "{Off-lightcone Wilson-line operators in gradient flow}",
    eprint = "2312.05032",
    archivePrefix = "arXiv",
    primaryClass = "hep-ph",
    doi = "10.1007/JHEP06(2024)210",
    journal = "JHEP",
    volume = "06",
    pages = "210",
    year = "2024"
}

@article{phan2019composable,
  title={Composable Effects for Flexible and Accelerated Probabilistic Programming in NumPyro},
  author={Phan, Du and Pradhan, Neeraj and Jankowiak, Martin},
  journal={arXiv preprint arXiv:1912.11554},
  year={2019}
}

@article{bingham2019pyro,
  author    = {Eli Bingham and
               Jonathan P. Chen and
               Martin Jankowiak and
               Fritz Obermeyer and
               Neeraj Pradhan and
               Theofanis Karaletsos and
               Rohit Singh and
               Paul A. Szerlip and
               Paul Horsfall and
               Noah D. Goodman},
  title     = {Pyro: Deep Universal Probabilistic Programming},
  journal   = {J. Mach. Learn. Res.},
  volume    = {20},
  pages     = {28:1--28:6},
  year      = {2019},
  url       = {http://jmlr.org/papers/v20/18-403.html}
}

@article{pymc2023,
  title = {{PyMC}: A Modern and Comprehensive Probabilistic Programming Framework in {P}ython},
  author = {Oriol Abril-Pla and Virgile Andreani and Colin Carroll and Larry Dong and Christopher J. Fonnesbeck and Maxim Kochurov and Ravin Kumar and Junpeng Lao and Christian C. Luhmann and Osvaldo A. Martin and Michael Osthege and Ricardo Vieira and Thomas Wiecki and Robert Zinkov },
  journal = {{PeerJ} Computer Science},
  volume = {9},
  number = {e1516},
  doi = {10.7717/peerj-cs.1516},
  year = {2023}
}

@article{Pineda:1997bj,
    author = "Pineda, A. and Soto, J.",
    editor = "Narison, Stephan",
    title = "{Effective field theory for ultrasoft momenta in NRQCD and NRQED}",
    eprint = "hep-ph/9707481",
    archivePrefix = "arXiv",
    reportNumber = "UB-ECM-PF-97-17",
    doi = "10.1016/S0920-5632(97)01102-X",
    journal = "Nucl. Phys. B Proc. Suppl.",
    volume = "64",
    pages = "428--432",
    year = "1998"
}

@article{Brambilla:1999xf,
    author = "Brambilla, Nora and Pineda, Antonio and Soto, Joan and Vairo, Antonio",
    title = "{Potential NRQCD: An Effective theory for heavy quarkonium}",
    eprint = "hep-ph/9907240",
    archivePrefix = "arXiv",
    reportNumber = "CERN-TH-99-199, HEPHY-PUB-716-99, UB-ECM-PF-99-06, UWTHPH-1999-34, UB-ECM-PF-99-13",
    doi = "10.1016/S0550-3213(99)00693-8",
    journal = "Nucl. Phys. B",
    volume = "566",
    pages = "275",
    year = "2000"
}

@article{Brambilla:2022rjd,
    author = "Brambilla, Nora and Chung, Hee Sok and Vairo, Antonio and Wang, Xiang-Peng",
    title = "{Production and polarization of S-wave quarkonia in potential nonrelativistic QCD}",
    eprint = "2203.07778",
    archivePrefix = "arXiv",
    primaryClass = "hep-ph",
    reportNumber = "TUM-EFT 168/22",
    doi = "10.1103/PhysRevD.105.L111503",
    journal = "Phys. Rev. D",
    volume = "105",
    number = "11",
    pages = "L111503",
    year = "2022"
}

@article{Lai:2025tpw,
    author = "Lai, Wai Kin and Chung, Hee Sok",
    title = "{Hadroproduction data support tetraquark hypothesis for {\ensuremath{\chi}}c1(3872)}",
    eprint = "2505.06910",
    archivePrefix = "arXiv",
    primaryClass = "hep-ph",
    doi = "10.1103/lkff-d2ph",
    journal = "Phys. Rev. D",
    volume = "112",
    number = "5",
    pages = "054005",
    year = "2025"
}

@misc{Brambilla:2026ujo,
    author = "Brambilla, Nora and Butenschoen, Mathias and Hibler, Simon and Mohapatra, Abhishek and Vairo, Antonio and Wang, Xiangpeng",
    title = "{Inclusive hadroproduction of $\chi_{c1}(3872)$, $X_b$ and pentaquarks}",
    eprint = "2602.14916",
    archivePrefix = "arXiv",
    primaryClass = "hep-ph",
    reportNumber = "TUM-EFT 203/26",
    month = "2",
    year = "2026"
}

@article{Sommer:1993ce,
    author = "Sommer, R.",
    title = "{A New way to set the energy scale in lattice gauge theories and its applications to the static force and $\alpha_s$ in SU(2) Yang-Mills theory}",
    eprint = "hep-lat/9310022",
    archivePrefix = "arXiv",
    reportNumber = "DESY-93-062",
    doi = "10.1016/0550-3213(94)90473-1",
    journal = "Nucl. Phys. B",
    volume = "411",
    pages = "839--854",
    year = "1994"
}

@article{DallaBrida:2019wur,
    author = "Dalla Brida, Mattia and Ramos, Alberto",
    title = "{The gradient flow coupling at high-energy and the scale of SU(3) Yang{\textendash}Mills theory}",
    eprint = "1905.05147",
    archivePrefix = "arXiv",
    primaryClass = "hep-lat",
    doi = "10.1140/epjc/s10052-019-7228-z",
    journal = "Eur. Phys. J. C",
    volume = "79",
    number = "8",
    pages = "720",
    year = "2019"
}

@article{Herren:2017osy,
    author = "Herren, Florian and Steinhauser, Matthias",
    title = "{Version 3 of RunDec and CRunDec}",
    eprint = "1703.03751",
    archivePrefix = "arXiv",
    primaryClass = "hep-ph",
    reportNumber = "TTP17-011",
    doi = "10.1016/j.cpc.2017.11.014",
    journal = "Comput. Phys. Commun.",
    volume = "224",
    pages = "333--345",
    year = "2018"
}

@article{ParticleDataGroup:2026aaa,
    author = "Takahashi, F. and others",
    collaboration = "Particle Data Group",
    title = "{Review of Particle Physics}",
    doi = "10.1142/S0217751X26300115",
    journal = "Int. J. Mod. Phys. A",
    volume = "41",
    pages = "2630011",
    year = "2026"
}

\clearpage
\onecolumngrid

\setcounter{section}{0}
\setcounter{subsection}{0}
\renewcommand{\thesection}{SM\arabic{section}}

\setcounter{secnumdepth}{2}

\renewcommand{\theequation}{SM\arabic{equation}}
\setcounter{equation}{0}
\onecolumngrid
\begin{center}
\large\textbf{Supplemental Material for\\ \enquote{Inclusive P-wave Quarkonium Decay Widths from Lattice QCD and pNRQCD}}

\end{center}
\vspace{1cm}
\section{Lattice calculation}
\label{sec:lattice_details}

We compute the chromoelectric correlator in Eq.~\eqref{eq:GEE_corr} on the lattice by expressing it in terms of bare link variables $U_\mu(x)$. To regulate the power divergence arising from the Wilson line and define a renormalized chromoelectric correlator, we implement the gradient flow method~\cite{Narayanan:2006rf,Luscher:2009eq,Luscher:2010iy} which introduces an auxiliary scale $\tau_F$, the flow time. Thus, at finite flow time, the bare link variables are replaced by their flowed counterparts.

The gradient flow renormalizes the chromoelectric fields and regulates the power divergence. We obtain an intermediate power-divergence-subtracted correlator by subtracting the leading flow-time-dependent power-divergence from the flowed correlator,
\begin{align}
    G_{EE}^{\rm sub}(\tau_F,t) = e^{\frac{d}{\sqrt{8\tau_F}} t}G_{EE}(\tau_F,t).\label{eq:GEE_sub}
\end{align}
The coefficient $d$, which parametrizes this divergence, is determined numerically to be $d=0.315(6)$, as described in Sec.~\ref{subsec:power_divergence}.

The organization of this section is as follows. We first describe the lattice setup and the main analysis strategy, including the continuum extrapolation at positive flow time, the zero-flow-time extrapolation of $G_{EE}^{\rm sub}(\tau_F,t)$, and the matching to the $\msbar$ correlator. The more technical ingredients are collected at the end of the section: the determination of the coefficient of the power divergence is described in Sec.~\ref{subsec:power_divergence}, while the details of the lattice tree-level calculation used for tree-level improvement are given in Sec.~\ref{subsec:lattice_tree_level}.

\subsection{Lattice setup}
\label{subsec:lattice_setup}

The lattice discretizations of the chromoelectric field components $E_{i,a}^{\rm lat}(t)$ and the lattice Wilson line $\Phi_{ab}^{\rm lat}(t,0)$ in the chromoelectric correlator~\eqref{eq:GEE_corr} are defined as
\be
    E_{i,a}^{\rm lat}(t)
    =
    \Tr\left[
    \lambda_a E_i^{\rm lat}(t)
    \right],
    \label{eq:lattice_E_adjoint}
\ee
and
\be
    \Phi_{ab}^{\rm lat}(t,0)
    =
    \frac{1}{2}
    \Tr\left[
    U(t,0)\lambda_a U(0,t)\lambda_b
    \right] .
    \label{eq:lattice_Phi_adjoint}
\ee
where $E_i^{\rm lat}(t)$ is the chromoelectric field in the fundamental representation and $U(0,t)$ is the fundamental Wilson line from the time slice at $0$ to that at $t$, and $U(t,0)=U^\dagger(0,t)$. The matrices $\lambda_a$ are the Gell-Mann matrices, normalized such that the $SU(3)$ generators are $T_a=\lambda_a/2$.

Inserting Eqs.~\eqref{eq:lattice_E_adjoint} and \eqref{eq:lattice_Phi_adjoint} into Eq.~\eqref{eq:GEE_corr}, and using the Fierz identity for the Gell-Mann matrices, gives the equivalent expression
\be
    G_{EE}^{\rm lat}(t)
    =
    -2
    \sum_{i=1}^3
    \left(
    \left\langle
    \Tr\left[
    gE_i^{\rm lat}(t) U(t,0)
    gE_i^{\rm lat}(0) U(0,t)
    \right]
    \right\rangle
    -
    \frac{1}{3}
    \left\langle
    \Tr\left[
    gE_i^{\rm lat}(t)
    \right]
    \Tr\left[
    gE_i^{\rm lat}(0)
    \right]
    \right\rangle
    \right) .
    \label{eq:lattice_EE_fund_app}
\ee
This form expresses the correlator entirely in terms of objects in the fundamental representation and is the one used in the numerical calculation.

For the chromoelectric field we use the two-plaquette discretization, denoted by 2Plaq. With this choice, the field insertions at the two ends of the temporal Wilson line are given by
\be
    gE_i^{\rm lat}(t)
    =
    \frac{i}{4}
    \left(
    U_{i,4}
    +
    U_{4,-i}
    -
    {\rm h.c.}
    \right),
    \label{eq:2plaq_definition_left}
\ee
and
\be
    gE_i^{\rm lat}(0)
    =
    \frac{i}{4}
    \left(
    U_{-i,-4}
    +
    U_{-4,i}
    -
    {\rm h.c.}
    \right).
    \label{eq:2plaq_definition_right}
\ee
Here $U_{\mu,\nu}$ denotes the untraced, oriented plaquette in the $(\mu,\nu)$ plane, starting and ending at the site of the field insertion, and ${\rm h.c.}$ denotes the Hermitian conjugate. In addition, we project the field components onto the traceless part by replacing
\be
    gE_i^{\rm lat}
    \longrightarrow
    gE_i^{\rm lat}
    -
    \frac{1}{3}
    \Tr\left[
    gE_i^{\rm lat}
    \right] .
    \label{eq:traceless_projection_E}
\ee
This projection corresponds to an $\mathcal{O}(a^2)$ improvement~\cite{Bilson-Thompson:2002xlt}. The 2Plaq discretization was found to perform better than the full clover-leaf discretization in the study of Ref.~\cite{Brambilla:2025cqy}.

The correlator defined above has mass dimension four and therefore scales as $a^{-4}$ in lattice units. In the analysis below we apply a tree-level improvement by normalizing the lattice correlator with its lattice tree-level expression and multiplying by the corresponding continuum tree-level correlator. This reduces leading discretization effects before the continuum extrapolation. The explicit tree-level calculation for the 2Plaq discretization is given in Sec.~\ref{subsec:lattice_tree_level}.

\subsection{Continuum limit, zero-flow-time extrapolation, and matching}
\label{subsec:lattice_analysis}

The three steps summarized in the paragraph around Eq.~\eqref{eq:matching}, and additionally the technical details and the tree-level improvement, are implemented as follows.

\subsubsection{Lattice ensembles and data sets}

\begin{table}[]
    \centering
    \caption{The simulation parameters used for the chromoelectric correlator.}
    \begin{tabular}{c|c|c|c|c|c}
        \hline\hline
        $N_S$ &  $n_t$ & $\beta$ & $t_0/a^2$ & $N_\mathrm{conf}$ & Label \\\hline
        26 & 52 & 6.481 & 13.44 & 6000 & L26 \\
        30 & 60 & 6.594 & 18.11 & 5831 & L30 \\
        40 & 80 & 6.816 & 32.18 & 3120 & L40 \\
        100 & 100 & 7.044 & 57.0 & 120 & L100 \\\hline
    \end{tabular}
    \label{tab:lattice_simulation_parameters}
\end{table}
Table~\ref{tab:lattice_simulation_parameters} summarizes the ensembles used in the analysis. Our primary data set consists of the L26, L30, and L40 ensembles. These ensembles are used for the main continuum extrapolation over the Euclidean-time range where all three lattice spacings provide reliable data.

To extend the analysis to smaller Euclidean times, we also use the L100 ensemble. 
At this very fine lattice spacing we observe reduced sampling of the topological sectors. We therefore use this ensemble only in the short-distance region, where sensitivity to finite-volume and long-distance topological effects is expected to be suppressed. We have explicitly checked the correlation of the short-distance observable with the sampled topological sectors and find no statistically significant dependence. This indicates that, in the range used in the analysis, the result is effectively insensitive to the particular topological sector sampled. This gives a second continuum data set, based on L30, L40, and L100, which allows us to reach smaller values of $t$.

In the following, we refer to the combination of the L26, L30, and L40 ensembles as the {\it lattice 1} data set. This data set is used over the main Euclidean-time range, where all three lattice spacings provide reliable data. We refer to the combination of the L30, L40, and L100 ensembles as the {\it lattice 2} data set. This second data set is used only in the short-distance region, where the inclusion of the L100 ensemble provides access to smaller Euclidean times.

\subsubsection{Tree-level improvement and interpolation}

Before taking the continuum limit, we apply a tree-level improvement to reduce discretization effects in the lattice correlator. We first form a dimensionless ratio by dividing the measured lattice correlator by the corresponding lattice tree-level correlator, computed with the same chromoelectric-field discretization. We perform the continuum extrapolation of this ratio and then restore the physical normalization by multiplying the continuum-extrapolated result by the continuum tree-level correlator. 
This correction removes the leading tree-level cutoff effects. The details of the lattice tree-level calculation, including the explicit expression for $G_{EE}^{\rm tree,lat}(a,\tau_F,t)$, are given in Sec.~\ref{subsec:lattice_tree_level}.

Before the continuum limit is taken, the lattice data are interpolated to common values of the dimensionless flow-time variable $\tau_F/t_0$. At fixed $\tau_F/t_0$, the correlator is then interpolated in the Euclidean time $t$ in units of $\sqrt{8 t_0}$. For most of the Euclidean-time range, we use cubic splines. At very small $t$, where the data can vary rapidly and the continuum extrapolation becomes more sensitive to local fluctuations, we use smoothing splines.

\subsubsection{Continuum limit at fixed positive flow time}

\begin{figure}
    \centering
    \includegraphics[width=0.48\linewidth]{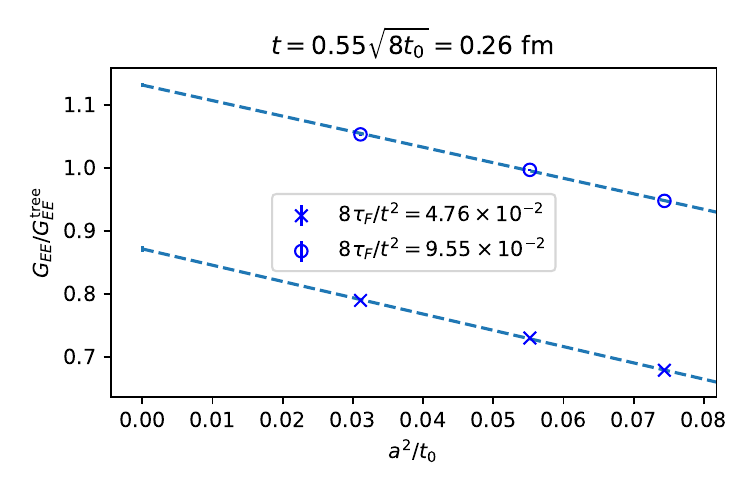}
    \includegraphics[width=0.48\linewidth]{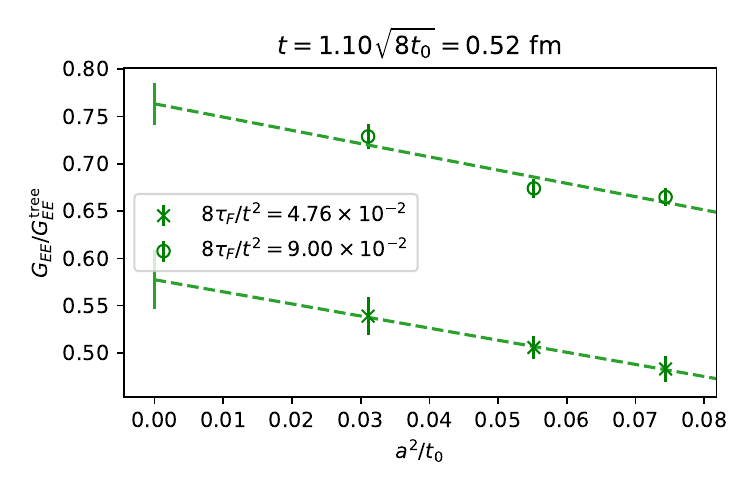}
    \caption{Examples of the continuum extrapolation of the tree-level improved correlator.
    The two panels correspond to two different Euclidean times $t$.
    In each panel, we show the continuum extrapolation at two different values of the flow time $\tau_F$.
    The plotted quantity is the lattice correlator normalized by the corresponding lattice tree-level correlator, as described in the text.
    }
    \label{fig:Continuum_limit_examples}
\end{figure}

The gradient flow provides a finite, regularized correlator at positive flow time. We therefore first take the continuum limit at fixed values of the dimensionless variables $t/\sqrt{t_0}$, and 
$\tau_F/t_0$. As described above, the extrapolation is performed on the tree-level normalized correlator, denoted by $G_{EE}/G_{EE}^{\rm tree}$ in Fig.~\ref{fig:Continuum_limit_examples}, using a linear fit in $a^2$. After the continuum limit of the ratio has been taken, we multiply by the continuum tree-level correlator to restore the physical normalization.

For the main range of Euclidean times we use the data {\it lattice 1}, consisting of L26, L30, and L40. For the smallest Euclidean times we use the data {\it lattice 2}, consisting of L30, L40, and L100. Figure~\ref{fig:Continuum_limit_examples} shows representative examples of the continuum extrapolation at two Euclidean-time separations and two different flow times.

\subsubsection{Subtraction of the flow-time dependent divergence}

At finite flow time, we subtract the power divergence by constructing $G_{EE}^{\rm sub}(\tau_F,t)$ in Eq.~\eqref{eq:GEE_sub} with $d=0.315(6)$. The numerical determination of $d$ is described in Sec .~\ref {subsec:power_divergence}. After this subtraction, and still at finite $\tau_F$, the finite matching factor of the chromoelectric correlator is unity up to corrections of order $\alpha_s^2$. The corresponding uncertainty is estimated and included in the final error budget, as described in the subsubsection entitled "{\it{Determination of $ \mathcal{E}_3^{\mathrm{np}}(t^\ast)$ and perturbative uncertainty}}" in Sec.~\ref{subsubsec:pert_unc}. We then take the zero-flow-time limit and denote the resulting correlator as
\be
    G_{EE}^{\rm sub}(t)
    =
    \lim_{\tau_F\to 0}
    G_{EE}^{\rm sub}(\tau_F,t) .
    \label{eq:subtracted_zero_flow_time_correlator}
\ee

\subsubsection{Zero-flow-time extrapolation}

After the continuum limit has been performed at fixed positive flow time, we extrapolate the subtracted correlator to $\tau_F\to0$. For each Euclidean time $t$, the extrapolation is performed using continuum-extrapolated data at several values of the flow radius $\sqrt{8\tau_F}$. The range of flow times included in the extrapolation is selected by imposing two conditions. First, the flow radius must be small compared with the Euclidean time separation, so that the zero-flow-time extrapolation remains controlled. Second, it must be sufficiently larger than the lattice spacing, so that the continuum limit at fixed positive flow time is reliable.
\begin{wrapfigure}{r}{0.42\textwidth}
    \centering
    \vspace{-0.5cm}
    \includegraphics[width=0.40\textwidth]{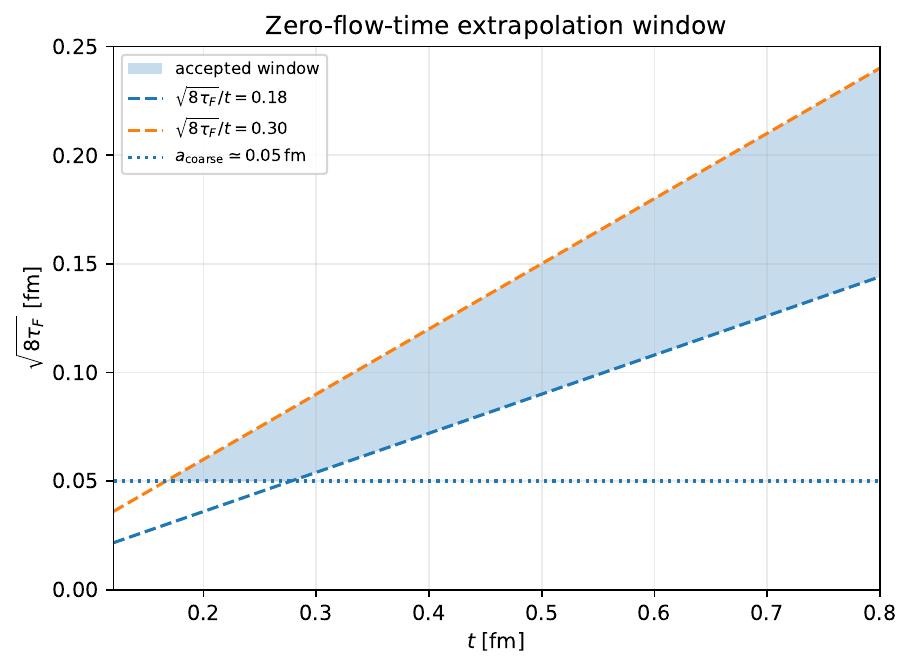}
    \caption{
    Window used for the zero-flow-time extrapolation in the $(t,\sqrt{8\tau_F})$ plane.
    The shaded region indicates the flow-time range included in the extrapolation.
    The diagonal lines correspond to the bounds
    $0.18 < \sqrt{8\tau_F}/t < 0.3$, while the horizontal line indicates the additional requirement
    $\sqrt{8\tau_F}\gtrsim a_{\rm coarse}\simeq \SI{0.05}{fm}$.
    }
    \label{fig:zftl_window_2d}
    \vspace{-0.3cm}
\end{wrapfigure}
In practice, we find stable results for
\be
    0.18 < \frac{\sqrt{8\tau_F}}{t} < 0.3 .
    \label{eq:zero_flow_time_window}
\ee
In addition, we require the flow radius to be larger than the coarsest lattice spacing used in the corresponding continuum extrapolation,
\be
    \sqrt{8\tau_F} \gtrsim a_{\rm coarse}
    \simeq \SI{0.05}{fm}.
    \label{eq:flow_radius_lattice_spacing_constraint}
\ee
This two-dimensional representation is shown in Fig.~\ref{fig:zftl_window_2d}. The diagonal lines correspond to the bounds in Eq.~\eqref{eq:zero_flow_time_window}, while the horizontal line indicates the additional requirement in Eq.~\eqref{eq:flow_radius_lattice_spacing_constraint}.

\begin{figure}
    \centering
    \includegraphics[width=0.48\linewidth]{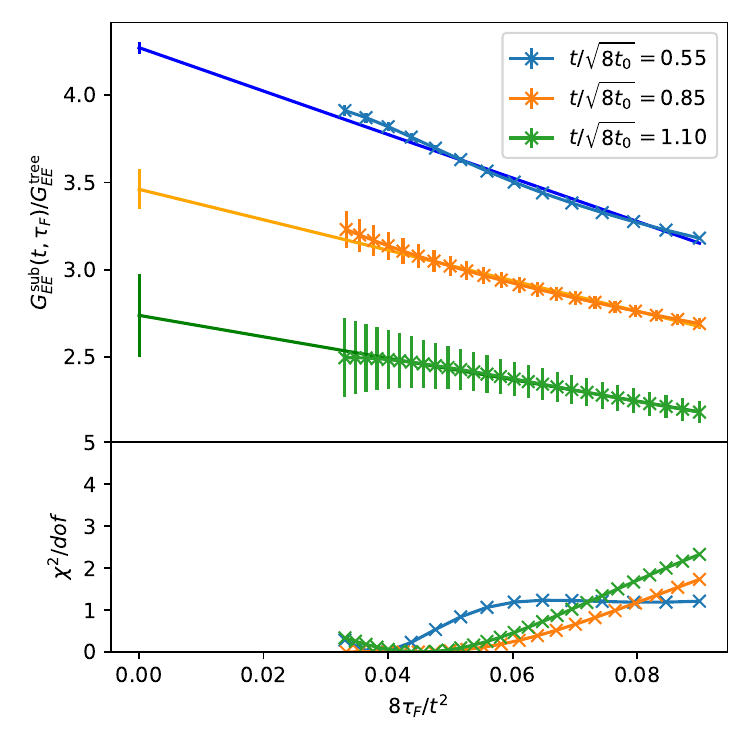}
    \includegraphics[width=0.48\linewidth]{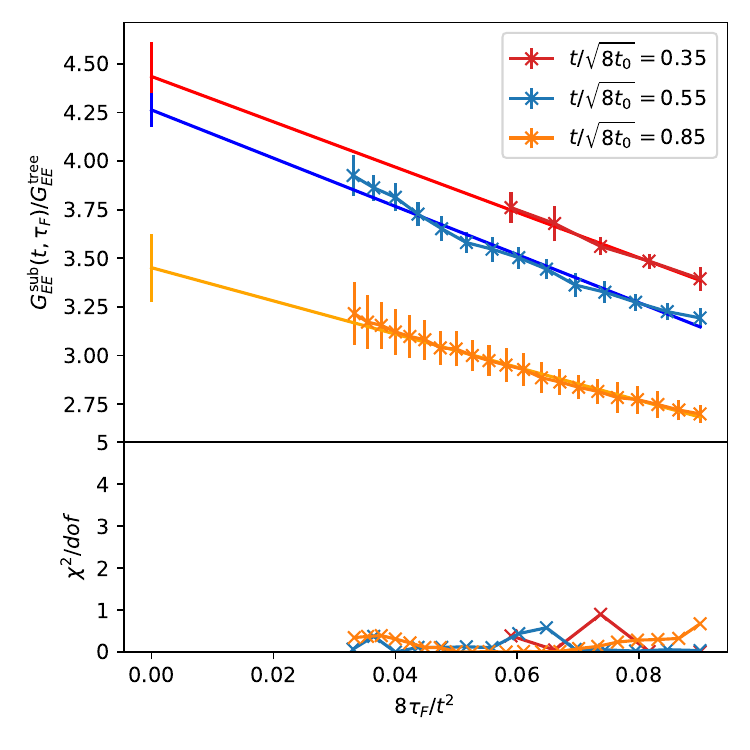}
    \caption{Examples of the zero-flow-time extrapolation for selected values of the Euclidean time $t$.
    The left panel uses continuum-extrapolated data obtained from the data {\it lattice 1}, with standard cubic splines used for the interpolation in the continuum-limit analysis.
    The right panel uses continuum-extrapolated data obtained from the data {\it lattice 2}, with smoothing splines used in the short-distance interpolation.
    The lower strip in each panel shows the corresponding values of $\chi^2_{\rm cont}/{\rm dof}$ from the continuum extrapolation at the flow times entering the zero-flow-time fit.
    }
    \label{fig:continuum_chi2_flowtime}
\end{figure}

Examples of the extrapolation to $\tau_F \to 0$ for selected values of $t$ are shown in Fig.~\ref{fig:continuum_chi2_flowtime}. The left panel uses the continuum-extrapolated data from data {\it lattice 1}, while the right panel uses the corresponding data from data {\it lattice 2}. The lower strip in each panel shows the values of $\chi^2_{\rm cont}/{\rm dof}$ for the continuum fits at the corresponding flow times entering the zero-flow-time extrapolation.

The quality of the continuum extrapolation is monitored for each flow time entering the zero-flow-time extrapolation. We include only data points for which the continuum fit satisfies the quality criterion $\chi^2_{\rm cont}/{\rm dof} < 3.2$. Within the window in Eq.~\eqref{eq:zero_flow_time_window}, the data are described by a linear extrapolation in the flow-time variable. We use the variation over acceptable fit ranges to monitor residual flow-time systematics.

\subsubsection{Matching to the $\msbar$ correlator}

The finite correlator $G_{EE}^{\rm sub}(t)$ is scheme dependent. A change of scheme corresponds to a multiplicative factor of the form
\be
    G_{EE}^{\msbar}(t)
    =
    e^{m_0^{\msbar}t}
    G_{EE}^{\rm sub}(t) ,
    \label{eq:msbar_matching_factor}
\ee
where $m_0^{\msbar}$ is the finite matching constant relating our intermediate scheme to the $\msbar$ scheme.

We determine $m_0^{\msbar}$ by matching the lattice correlator to the NLO perturbative expression of $\mathcal{E}(t)$ in $\msbar$ at short Euclidean time. The matching is performed for
\be
    t_{\rm min} \leq t \leq t_{\rm max},
    \qquad
    t_{\rm max} = 0.65\sqrt{8t_0} = \SI{0.30}{fm}.
    \label{eq:matching_tmax}
\ee
We have varied both the values of $t_{\rm min}$ and $t_{\rm max}$ within reason without noticing any appreciable difference.
The perturbative expression is evaluated at the scale
\be
    \mu = \SI{1.17}{\GeV}
    = \frac{2e^{-\gamma_E}}{t_\ast},
    \label{eq:matching_scale}
\ee
with $\alpha_s^{(n_f=0)}(\mu)=0.26$ at two loops determined using RunDec~\cite{Chetyrkin:2000yt,Herren:2017osy}.

We perform the matching separately for the two continuum limits using both data sets {\it lattice 1} and {\it lattice 2}. The lower end of the matching window, $t_{\rm min}$, is set by the smallest Euclidean time reached by each data set after the continuum and zero-flow-time extrapolations. For the data set {\it lattice 1}, the lower end of the matching window is
\be
    t_{\rm min}^{(1)} = \SI{0.236}{fm}
    = 0.50\sqrt{8t_0}.
    \label{eq:smallest_time_dataset1}
\ee
This data set gives a less satisfactory short-distance matching, with $\chi^2/{\rm dof}=15.1$. The data set {\it lattice 2} reaches the smaller Euclidean time
\be
    t_{\rm min}^{(2)} = \SI{0.165}{fm}
    = 0.35\sqrt{8t_0},
    \label{eq:smallest_time_dataset2}
\ee
and gives a good fit quality, $\chi^2/{\rm dof}=0.1$. The two determinations are combined in a weighted average, with the data set {\it lattice 2} receiving the larger weight because it reaches shorter Euclidean times and gives a better matching fit.

The resulting estimate is
\be
    m_0^{\msbar}\sqrt{8t_0} = 0.476(12),
    \label{eq:m0_msbar_result}
\ee
where the quoted uncertainty is statistical. It is obtained from the jackknife samples used to propagate the uncertainty through the continuum and zero-flow-time extrapolations. The resulting jackknife samples of $m_0^{\msbar}$ are then used to construct the corresponding samples for $\mathcal{E}_3^\mathrm{np}(t_\ast)$.

\subsubsection{Determination of $\mathcal E_3^{\text{np}}(t_\ast)$ and perturbative uncertainty}
\label{subsubsec:pert_unc}

We calculate $\mathcal{E}_3^{\mathrm{np}}$ using the matched correlator $G_{EE}^{\msbar}(t)$ in Eq.~\eqref{eq:E3_lattice_part}. The integrand is smooth in $t$ as Fig.~\ref{fig:Gr_msBar_matching} shows the smooth transition from the perturbative expression to the lattice correlator. Consequently, the residual dependence on the separation scale $t ^\ast$ is mild and is dominated by statistical fluctuations.

To estimate the uncertainty associated with missing higher orders in the perturbative expression used to determine the finite matching factor, we repeat the analysis after multiplying the perturbative correlator by $1 \pm \alpha_s^2$. This changes the fitted value of $m_0^{\msbar}$ and, consequently, that of $\mathcal E_3$. The induced variation is used as an estimate of the systematic uncertainty associated with missing higher-order perturbative corrections and is included in the final error budget (see Eq.~\eqref{eq: E3choice}).

\subsection{Determination of the power divergence}
\label{subsec:power_divergence}

\begin{table}[]
\centering
\begin{tabular}{ll|ll|ll|ll|ll}
\multicolumn{2}{l|}{$\beta=6.481$} & \multicolumn{2}{l|}{$\beta=6.510$} & \multicolumn{2}{l|}{$\beta=6.594$} & \multicolumn{2}{l|}{$\beta=6.700$} & \multicolumn{2}{l}{$\beta=6.816$} \\
\multicolumn{2}{l|}{$N_s=26$}      & \multicolumn{2}{l|}{$N_s=30$}     & \multicolumn{2}{l|}{$N_s=30$}      & \multicolumn{2}{l|}{$N_s=40$}      & \multicolumn{2}{l}{$N_s=40$}      \\ \hline
$n_t$           & $T/T_C$          & $n_t$           & $T/T_C$          & $n_t$           & $T/T_C$          & $n_t$           & $T/T_C$          & $n_t$          & $T/T_C$          \\ \hline
4               & 3.65             & 4               & 3.80             & 4               & 4.24             & 4               & 4.87             & 4              & 5.65             \\
6               & 2.43             & 6               & 2.53             & 6               & 2.83             & 6               & 3.25             & 6              & 3.77             \\
8               & 1.83             & 8               & 1.90             & 8               & 2.12             & 8               & 2.44             & 8              & 2.82             \\
10              & 1.46             & 10              & 1.52             & 10              & 1.70             & 10              & 1.95             & 10             & 2.26             \\
12              & 1.22             & 12              & 1.27             & 12              & 1.41             & 12              & 1.62             & 12             & 1.89             \\
14              & 1.04             & 14              & 1.08             & 14              & 1.21             & 14              & 1.39             & 14             & 1.61             \\
                &                  &                 &                  & 16              & 1.06             & 16              & 1.22             & 16             & 1.41             \\
                &                  &                 &                  &                 &                  & 18              & 1.08             & 18             & 1.26             \\
                &                  &                 &                  &                 &                  &                 &                  & 20             & 1.13             \\
                &                  &                 &                  &                 &                  &                 &                  & 22             & 1.03            
\end{tabular}
\caption{The simulation parameters for additional finite $T$ ensembles to extract the divergence from the Polyakov loop. For each parameter set, 1000 configurations were sampled. One column corresponds to a fixed $\beta$ and spatial lattice extent, i.e., the same lattice spacing and fixed spatial volume.}
\label{tab:finite_T_divergence_ensembles}
\end{table}

In a cutoff regulator scheme, such as a lattice regulator or a gradient flow regulator, certain correlators exhibit a power divergence. This divergence is associated with Wilson lines of a given length. For instance, the chromoelectric correlator includes a temporal Wilson line of length $t$. At finite gradient flow time, the divergent contribution per unit length takes the form $d/\sqrt{8\tau_F}$. Therefore, studying correlators as a function of flow time provides insights into this divergent behavior, allowing us to extract the value of $d$. However, determining the power-divergent contribution directly from $G_{EE}$ would require substantially higher statistical precision. We therefore determine $d$ from a different observable with the same power divergence.

At finite temperature, the Polyakov loop in the adjoint representation,
\begin{align}
    P_8 \equiv \Phi_{aa}(1/T,0)
\end{align}
where $T$ is the temperature, is a Wilson line with length equal to the temporal extent of the lattice. Consequently, it carries the same power divergence per unit length as the chromoelectric correlator. The expectation value of the Polyakov loop $L_8 = \frac{1}{8} \langle P_8\rangle$ has the divergent structure
\begin{align}
    L_8 = e^{-\frac{d}{\sqrt{8\tau_F}T}}L_8^{\text{sub}}
\end{align}
where $L_8^{\text{sub}}$ denotes the Polyakov loop after subtraction of the leading power-divergent contribution. 
Unlike the chromoelectric correlator, the Polyakov loop can be computed with high precision. We conduct additional finite-temperature calculations across five different lattice spacings, each with a set of different temporal extents, covering a wide temperature range. The simulation parameters are listed in table~\ref{tab:finite_T_divergence_ensembles}.

To extract the parameter $d$, we perform a Bayesian fit to the flowed Polyakov loops, at fixed lattice spacing. We fit the logarithm of the expectation value of the Polyakov loop and model the remaining flow time dependence using polynomials of degree $n$. The fit function for $\log L_8$ thus takes the form
\begin{align}
f_n(\tau_F) = C_0 - \frac{d}{\sqrt{8\tau_F}T} + m_0 \tau_F T^2 + P_n(\sqrt{8\tau_F}),
\end{align}
where $C_0$ and $m_0$ are additional fit parameters, and $P_n(x)$ is a polynomial of degree $n$. This polynomial is defined by a set of coefficients $a_i^{(n)} \equiv \mathbf{a}^{(n)}$ for $i = 1, 3, \ldots, n$ (with $0$ and $2$ excluded since they are accounted for by $C_0$ and $m_0$).

We assume that the measured data is normally distributed around the mean value $f_n(\tau_F)$, that is, $\log L_8 \sim \mathcal{N}(f_n(\tau_F), \sigma_{\tau_F})$. For the priors of $C_0$ and $d$, denoted as $p(C_0)$ and $p(d)$, we assume a uniform distribution with bounds that are far from the maximum likelihood estimation (MLE), effectively simulating a flat prior while ensuring numerical stability. For the parameter $m_0$, we take a normal distribution with a mean of $0$ and a standard deviation of $\sigma = 1$, indicating a relatively broad prior in comparison to the expected parameter range.

For the coefficients $a_i^{(n)}$, we assume a normal prior distribution $p(\{a_i^{(n)}\})$, where $\{...\}$ denotes the collection of coefficients, with a mean $\mu_{\mathbf{a}^{(n)}} = 0$ and a standard deviation\footnote{We consider $\sigma_{\mathbf a^{(n)}}=1$ and $3$ in order to test the sensitivity of the fit to the width of the prior distribution.} $\sigma_{\mathbf{a}^{(n)}} = 1$ or $\sigma_{\mathbf{a}^{(n)}} = 3$. For the standard deviation of the data, $p(\{\sigma_{\tau_F}\})$, we assume a weakly informative inverse gamma distribution with parameters $\alpha = 3$ and $\beta = \frac{1}{2}$. $\{\sigma_{\tau_F}\}$ is the set of the standard deviations of all data points for each flow time.

In summary, for a given model function $f_n$, the likelihood for a measured value of $\log L_8$ is
\begin{align}
p(\log L_8 \mid C_0,d,m_0,\mathbf{a}^{(n)},\{\sigma_{\tau_F}\},f_n)
=
\mathcal{N}\!\left(\log L_8 \mid f_n(\tau_F),\{\sigma_{\tau_F}\}\right).
\end{align}
The corresponding posterior distribution for the parameters and the model is then given, up to an overall normalization, by
\begin{align}
&p(C_0,d,m_0,\mathbf{a}^{(n)},\{\sigma_{\tau_F}\},f_n \mid \log L_8) \nonumber\\
&\qquad\propto
\mathcal{N}\!\left(\log L_8 \mid f_n(\tau_F),\{\sigma_{\tau_F}\}\right)
p(C_0) p(d) p(m_0) p(\mathbf{a}^{(n)}) p(\{\sigma_{\tau_F}\}) p(f_n).
\end{align}
Consequently, for a data set $(\log L_8)^j$
where $j$ labels the data points and $N$ is the total number of data points, the likelihood is
\begin{align}
p(\{(\log L_8)^j\}_{j=1}^N \mid C_0,d,m_0,\mathbf{a}^{(n)},\{\sigma_{\tau_F}\},f_n) =
\prod_{j=1}^N p((\log L_8)^j \mid C_0,d,m_0,\mathbf{a}^{(n)},\{\sigma_{\tau_F}\},f_n).
\end{align}
For reasons of efficiency and stability, we use the data as a blocked jackknife sample with 30 blocks.

In this model, the remaining flow-time dependence of the Polyakov loop after subtraction of the leading power divergence is modeled by the polynomial of degree $n$ in the functional form $f_n$. There is no designated value for $n$, and to account for this uncertainty, we build a final model that performs a weighted sum of different values of $n$ with weights $w_n$ and a Dirichlet distribution with $\alpha_n=2$ for the prior $p(\{w_n\}_{n=n_1}^{n_2})$. This model averaging follows the same logic as information-criterion-based model averaging, such as the Widely Applicable Information Criterion (WAIC).

The final likelihood therefore reads
\begin{align}
    p(\{(\log L_8)^j\}_{j=1}^N|\Theta,M)
    &=\sum_{n=n_1}^{n_2} w_n \prod_{j=1}^N \mathcal{N}\!\left((\log L_8)^j\mid f_n(\tau_F^j),\sigma_{\tau_F^j}\right),
    \label{eq:bayesian_fit_function}
\end{align}
where $\Theta=(C_0,d,m_0,\{\mathbf{a}^{(n)}\},\{\sigma_{\tau_F}\},\{w_n\}_{n=n_1}^{n_2})$ denotes the set of model parameters, and $M$ collects the remaining hyperparameters that must be chosen. Explicitly, those are the priors for the coefficients $\mathbf{a}^{(n)}$, $n_1$, $n_2$, and the flow time range considered in the fit. We fix $n_1=3$ and $n_2=6$, which means we model the finite part with polynomials up to degree 6. We test $\sigma_{\mathbf{a}^{(n)}}=1$ and $\sigma_{\mathbf{a}^{(n)}}=3$.

\begin{figure}
    \centering
    \includegraphics[width=0.49\linewidth]{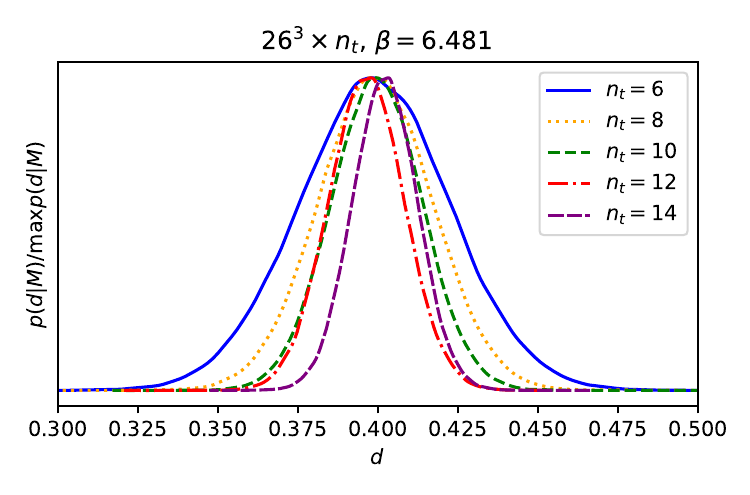}
    \includegraphics[width=0.49\linewidth]{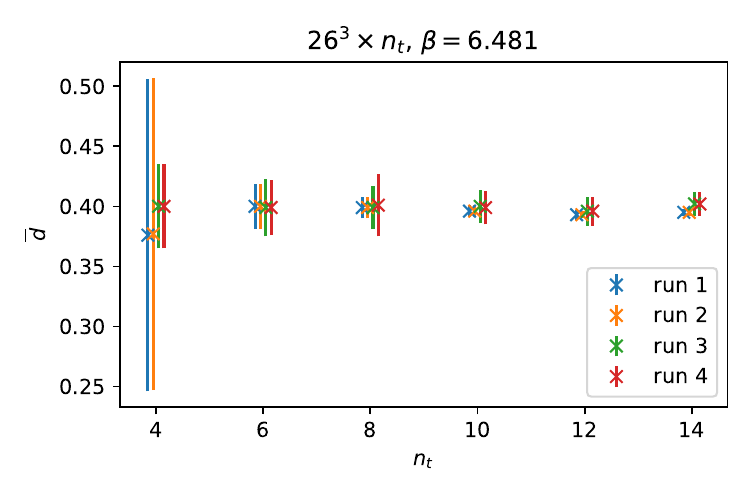}
    \caption{The left plot shows the functional form of the distribution for $d$ for different temporal lattice extents. The right plot shows the mean and standard deviation of $d$ across different lattice sizes for four sets of hyperparameters in the Bayesian sampling.}
    \label{fig:d_fit_L26}
\end{figure}

With the likelihood Eq.~\eqref{eq:bayesian_fit_function} and the observed data set $\{(\log L_8)^j\}_{j=1}^N$, we can apply Bayes theorem to find the distribution of the model parameters
\begin{align}
p(\Theta|\{(\log L_8)^j\}_{j=1}^N,M) = \frac{p(\{(\log L_8)^j\}_{j=1}^N|\Theta,M)p(\Theta|M)}
{p(\{(\log L_8)^j\}_{j=1}^N|M)}\,.
\label{eq:model_parameter_distribution}
\end{align}
The denominator does not depend on $\Theta$; therefore, for parameter inference, it only contributes to the normalization of the posterior distribution.

Using Eq.~\eqref{eq:model_parameter_distribution}, we can sample the parameters $C_0$, $d$, $m_0$, $\{\mathbf{a}^{(n)}\}$, $\{\sigma_{\tau_F}\}$, and $\{w_n\}_{n=n_1}^{n_2}$ with Hamiltonian Monte Carlo (HMC) methods. Our primary interest is the parameter $d$. By focusing on this parameter, we marginalize over the remaining parameters, leading to
\begin{align}
p(d | \{(\log L_8)^j\}_{j=1}^N,M) &= \int \mathrm{d}C_0\, \mathrm{d}m_0\,
\mathrm{d}\{\mathbf{a}^{(n)}\}\, \mathrm{d}\{\sigma_{\tau_F}\}\,
\mathrm{d}\{w_n\}_{n=n_1}^{n_2} \nonumber\\
&\quad \times p(C_0,d,m_0,\{\mathbf{a}^{(n)}\},\{\sigma_{\tau_F}\},\{w_n\}_{n=n_1}^{n_2} | \{(\log L_8)^j\}_{j=1}^N,M)\,.
\end{align}
We use PyMC~\cite{pymc2023} to build the Bayesian model and a NumPyro implementation~\cite{phan2019composable,bingham2019pyro} of the No-U-Turn Sampler (NUTS)~\cite{hoffman2014no} to sample from the posterior. NumPyro can be used as the backend sampler in PyMC and is an efficient HMC sampler. Figure~\ref{fig:d_fit_L26} illustrates the posterior distributions of the parameter $d$ and the stability of its extracted value across different temporal lattice extents and Bayesian hyperparameter choices for $\beta = 6.481$. In the left plot, we present the distribution of $d$ for different temporal lattice extents. We observe that the mean value remains constant across the different values of $n_t$, while the standard deviation decreases as $n_t$ increases. Therefore, we conclude that $d$ is independent of $n_t$. In addition, a larger $n_t$ stabilizes the extraction by providing a wider range of Euclidean times, allowing the analysis to extend to larger flow times.

\begin{wrapfigure}{r}{0.5\textwidth}
    \vspace{-0.5cm}
    \centering
    \includegraphics[width=1\linewidth]{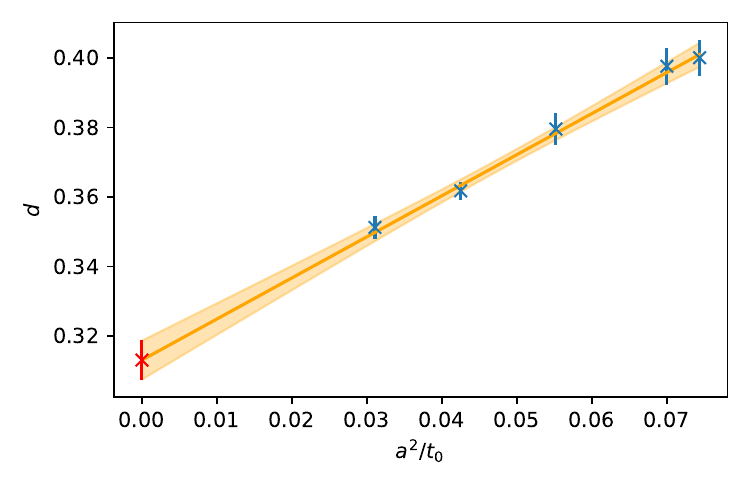}
    \caption{The continuum limit of the parameter $d$.}
    \label{fig:d_continuum_limit}
    \vspace{-0.3cm}
\end{wrapfigure}

The right plot of the same figure shows the mean values and their standard deviations for various lattice sizes and hyperparameter sets. ``run 1(3)" uses $\sigma_{\mathbf{a}^{(n)}} = 1$, while ``run 2(4)" employs $\sigma_{\mathbf{a}^{(n)}} = 3$. ``run 1(2)" uses a flow time range of $a < \sqrt{8\tau_F} < 0.42/T$, whereas ``run 3(4)" is fixed in the range $a < \sqrt{8\tau_F} < 2.33a$. The first range expands with increasing temporal lattice extent $n_t$, while the second remains constant across all temperatures.

From our analysis, we make two key observations: first, all runs give compatible mean values, indicating that our model is robust under the tested hyperparameter choices. Second, we observe no significant temperature dependence of $d$, as expected, since the power divergence is a UV divergence and thus depends only on the cutoff. We determine the final value for $d$ at each lattice spacing through a constant fit across all $n_t$.

We repeat the process for each lattice spacing, followed by a continuum extrapolation of $d$ using a linear fit in $a^2$. Figure~\ref{fig:d_continuum_limit} shows the continuum extrapolation. We use $t_0/a^2$ for $\beta=6.481,6.594,6.816$ from our zero temperature calculations, and for $\beta=6.510,6.700$ we use the scale setting in Ref.~\cite{Francis:2015lha}. The final continuum result is $d=0.315(6)$.

\subsection{Lattice tree-level calculation}
\label{subsec:lattice_tree_level}
We study the leading discretization artifacts through a tree-level calculation of the correlator. 
The link variables are written in terms of the gauge fields as $\displaystyle U_\mu(n)=e^{ig_0A_\mu^a(n)T_a}$ where $g_0$ is the bare coupling, and we set $a=1$. 
The Fourier transform of the gauge fields yields $\displaystyle  A_\mu^a(n)=\int_k e^{i(k\cdot n + k_\mu/2)}\tilde{A}_\mu(k)$ where $\tilde{A}_\mu(k)$ is the gauge field in momentum space. 
The gluon propagator in Feynman gauge is
\begin{align}
    \left\langle 0\left|
    \tilde{A}_\mu^a(k)\tilde{A}_\nu^b(k')
    \right| 0\right\rangle
    =
    (2\pi)^4\delta(k+k')\delta_{ab}\delta_{\mu\nu}\frac{1}{D(k)}
    =
    (2\pi)^4\delta(k+k')\delta_{ab}\delta_{\mu\nu}
    \frac{1}{4\sum_{i=1}^4\sin^2 (k_i/2)} .
\end{align}
Since $E_i\sim\mathcal{O}(g_0)$ and $U(0,t)\sim1+\mathcal{O}(g_0)$, only the $E$-fields contribute at tree level. Therefore, the tree-level result for the 2-plaquette discretization is
\begin{align}
    G_{EE}^\mathrm{tree}(\sqrt{8\tau_F},t) = g_0^2\int_{0}^\pi \frac{\mathrm{d}^4k}{(2\pi)^4} \ e^{-2\tau_FD_F(k)} \frac{128 \cos [k_4(t+1)]\sum_{i=1}^3(\cos k_i+1)(2-\cos k_4-\cos k_i)}{D(k)}
    \label{eq:2plaq_lattice_tree_level}
\end{align}
Here $D(k)=4\sum_{i=1}^4\sin^2 (k_i/2)$ is the Wilson gauge-action kernel, while $D_F(k)=D(k)+4\sum_{i=1}^4\sin^4 (k_i/2)/3$ is the tree-level Symanzik-flow kernel used to flow the fields. We have symmetrized the integrand so that the integration bounds are $k_\mu\in[0,\pi]$. In the numerical determination of the integral, we use the regulated range $k_\mu\in[\epsilon,\pi-\epsilon]$ and perform an $\epsilon\rightarrow0$ extrapolation.

\begin{figure}
    \centering
    \includegraphics[width=0.48\linewidth]{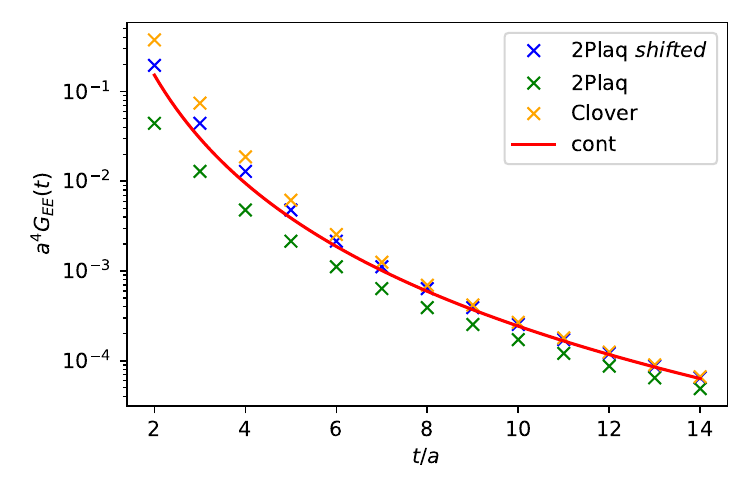}
    \caption{Comparison of the different tree-level results at zero flow time. Data points ``2Plaq" represent the correlator following definition of Eq.~\eqref{eq:lattice_EE_fund_app} with fields defined in Eqs.~\eqref{eq:2plaq_definition_left} and~\eqref{eq:2plaq_definition_right}. Data points ``2Plaq \emph{shifted}" represent the same function as ``2Plaq", but the effective time separation $t$ is shifted by one lattice spacing. ``Clover" is the tree level result of the correlator with clover leaf discretization of the $E$-fields. ``Cont" is the continuum tree level result.}
    \label{fig:latt_cont_tree_level_comparison}
\end{figure}

Figure~\ref{fig:latt_cont_tree_level_comparison} shows the tree-level result for various setups. We find that the 2Plaq discretization has the smallest discretization artifacts once the effective time separation $t$ is adjusted by one lattice spacing to account for the natural lattice location of the electric field operator.

\section{Details on $\mathcal{E}$, $\mathcal{E}^{\mathrm{pt}}_3$ and  Short Distance Coefficients}
The chromoelectric correlator regularized in $D$-dimensions reads~\cite{Eidemuller:1997bb, Braun:2020ymy}
\begin{align}
    \mathcal{E}\left(t^2\right)=-4\pi\alpha_{s}\left[\mathcal{E}^{(0)}(t^{2})+\mathcal{E}^{(1)}(t^{2})+t^{2}\frac{\partial}{\partial t^{2}}\mathcal{E}^{(1)}(t^{2})\right],
\label{eq:EMSNLO}
\end{align}
where at order $\mathcal{O}(\alpha_{s}^{2})$ (1 loop) for $D=4-2\varepsilon$,
\begin{align}
    \mathcal{E}^{(0)}(t^{2})&=N_{c}\left(N_{c}^{2}-1\right)\frac{\alpha_{s}}{4\pi}\frac{\Lambda^{4\varepsilon}}{\pi^{2-2\varepsilon}}\Gamma^{2}\left(1-\varepsilon\right)\frac{1}{\left(t^{2}\right)^{2-2\varepsilon}}g(\varepsilon),\\\mathcal{E}^{(1)}(t^{2})&=\left(N_{c}^{2}-1\right)\Lambda^{2\varepsilon}\frac{\Gamma\left(2-\varepsilon\right)}{\pi^{2-\varepsilon}\left(t^{2}\right)^{2-\varepsilon}}\left[1+N_{c}\frac{\alpha_{s}}{4\pi}\frac{\Lambda^{2\varepsilon}}{\pi^{-\varepsilon}}\frac{\Gamma^{2}\left(1-\varepsilon\right)}{\Gamma\left(2-\varepsilon\right)}\left(\frac{1}{t^{2}}\right)^{-\varepsilon}g_{1}(\varepsilon)\right],
\end{align}
with
\begin{align}
    g(\varepsilon)&=-\frac{1}{\varepsilon}+3+\mathcal{O}(\varepsilon),\qquad
g_1(\varepsilon)=\frac{1}{3N_c}\left[\frac{8N_c+4T_Fn_f}{\varepsilon}+\frac{1}{3}\left(\left(37+12\pi^2\right)N_c+4T_Fn_f\right)\right]+\mathcal{O}(\varepsilon).    
\end{align}  
At one loop, the ultraviolet divergence of the correlator is removed entirely through the renormalization of the strong coupling, which we perform in the $\msbar$ scheme. 

The quantity $\mathcal{E}_3$ is the sum of a perturbative and a nonperturbative component. 
Here we give the perturbative component, $\mathcal{E}_3^{\mathrm{pt}}$, at $\mathcal{O}(\alpha_s^2)$. 
In $\msbar$, it reads
\begin{align}\label{eq: E3nlo}
   \mathcal{E}^{\mathrm{pt}}_{3}\left(\Lambda,t_{\ast}\right)
   &=\frac{T_F}{N_c}\int^{t_\ast}_0\mathrm{d}t\,t^3\mathcal{E}^{\msbar}(t)\nonumber\\
   &=24C_{F}\,\frac{\alpha_s(\Lambda)}{4\pi}\left\{-\frac{8}{3}+ L_{t_\ast}+\frac{\alpha_s}{4\pi}\left[\frac{\beta_0}{2}L_{t_\ast}^{2}+\gamma^{(1)}L_{t_\ast}+c\right]\right\} +\mathcal{O}\left(\alpha_{s}^{3}\right),
\end{align}
where $\Lambda$ is the (soft) renormalization scale of the correlator, the renormalized coupling is in the $\msbar$ scheme, and $\displaystyle L_{t_\ast}=\log\left(\frac{e^{2\gamma_E}\Lambda^{2}t_{\ast}^{2}}{4}\right)$. 
Moreover, we have defined 
\begin{align}
    &\qquad\gamma^{(1)}=\frac{4}{9}\left[N_{c}\left(7+3\pi^{2}\right)+n_{f}T_{F}\right],\\
    c=-\frac{\beta_{0}}{2}l_{\msbar}^2+\frac{4}{9}&\left[7n_fT_F-N_c\left(29+3\pi^2\right)\right]l_{\overline{\rm MS}}
   +\frac{16}{9}\left[n_{f}T_{F}-N_{c}\left(2+2\pi^{2}+9\zeta(3)\right)\right],
\end{align}
with $\displaystyle l_{\msbar}\equiv\gamma_E-\ln4\pi$, $\beta_0=11N_c/3-4T_Fn_f/3$, $C_{F}=(N_{c}^{2}-1)/(2N_{c})$, $n_{f}$ the number of massless quark flavors and $T_{F}=1/2$. 
In Eq.~\eqref{eq: E3nlo}, $\Lambda$ plays also the role of the renormalization/factorization scale of the time integral in~$\mathcal{E}^{\mathrm{pt}}_3$. 

For the final evaluation of inclusive heavy quarkonium decays, we use the perturbative expansions of the SDCs at $\mathcal{O}(\alpha^3_s)$. They read in the $\msbar$ scheme \cite{Petrelli:1997ge},
\begin{align}\label{eq:SDCs} 
\mathrm{Im}f_{1}\left(^{3}P_{0}\right)=&\frac{3\pi C_{F}}{2N_{c}}\alpha_{s}^{2}\Biggl\{1+\frac{\alpha_{s}}{\pi}\biggl[C_{F}\left(-\frac{7}{3}+\frac{\pi^{2}}{4}\right)+C_{A}\left(\frac{427}{81}-\frac{\pi^{2}}{144}\right)\nonumber\\ 
&+\beta_{0}\log\left(\frac{\mu_{H}}{2m_Q}\right)\biggr]\Biggr\}+n_{f}\frac{2C_{F}}{9N_{c}}\alpha_{s}^{3}\left[-\frac{29}{6}-\mathrm{log}\left(\frac{\Lambda}{2m_Q}\right)\right],\\\nonumber\\
\mathrm{Im}f_{1}\left(^{3}P_{1}\right)=&\frac{C_{F}}{4}\alpha_{s}^{3}\left(\frac{587}{27}-\frac{317}{144}\pi^{2}\right)+n_{f}\frac{2C_{F}}{9N_{c}}\alpha_{s}^{3}\left[-\frac{4}{3}-\mathrm{log}\left(\frac{\Lambda}{2m_Q}\right)\right],\\\nonumber\\
\mathrm{Im}f_{1}\left(^{3}P_{2}\right)=&\frac{2\pi C_{F}}{5N_{c}}\alpha_{s}^{2}\Biggl\{1+\frac{\alpha_{s}}{\pi}\biggl[-2C_{F}+N_{c}\left(\frac{2185}{216}-\frac{337\pi^{2}}{384}+\frac{5}{3}\log\left(2\right)\right)\nonumber\\  
   &+\beta_{0}\log\left(\frac{\mu_{H}}{2m_Q}\right)\biggr]\Biggr\}+n_{f}\frac{2C_{F}}{9N_{c}}\alpha_{s}^{3}\left[-\frac{29}{15}-\mathrm{log}\left(\frac{\Lambda}{2m_Q}\right)\right],\\\nonumber\\
   \mathrm{Im}f_{8}\left(^{3}S_{1}\right)=&\frac{\pi n_{f}}{6}\alpha_{s}^{2}\Biggl\{1+\frac{\alpha_{s}}{\pi}\biggl[-\frac{13}{4}C_{F}+N_{c}\left(\frac{133}{18}-\frac{\pi^{2}}{4}+\frac{2}{3}\log\left(2\right)\right)\nonumber\\ 
   &-\frac{10}{9}n_{f}T_{F}+\beta_{0}\log\left(\frac{\mu_{H}}{2m_Q}\right)\biggr]\Biggr\}+\frac{5}{2}\alpha_{s}^{3}\left(-\frac{73}{4}+\frac{67\pi^{2}}{36}\right),
\end{align}
where $m_Q$ is the mass of the heavy quark of flavor $Q$ and $\mu_H$ is the (hard) renormalization scale. 
A natural scale choice is $\mu_H=m_Q$. 
We estimate the error from truncating the perturbative series of the SDCs by multiplying the $\mathcal{O}(\alpha_s^3)$ contribution to the SDC by $\alpha_s$.

\end{document}